\documentclass[useAMS,usenatbib]{mn2e}

\usepackage[T1]{fontenc}
\usepackage{aecompl}
\usepackage{amssymb}
\usepackage{aas_macros}
\usepackage{deluxetable}
\usepackage{natbib}
\usepackage{multirow}
\usepackage{hyperref}
\usepackage{color}
\usepackage{epsfig}
\usepackage{multicol}
\usepackage{lscape}
\usepackage{pdfcomment}
\usepackage{soul}
\usepackage{subfig,float}

\usepackage[english]{babel}
\usepackage[utf8x]{inputenc}
\usepackage[colorinlistoftodos]{todonotes}
\usepackage{marginnote}

\newcommand{\bhmass}{${\cal M}_{\rm BH}$} %{${\cal M}_\bullet$}
\newcommand{\cmomass}{${\cal M}_{\rm CMO}$}
\newcommand{\nscmass}{${\cal M}_{\rm NSC}$}
\newcommand{\nscsize}{$r_{\rm eff, NSC}$}
\newcommand{\smbhsize}{$r_{\rm infl, BH}$}
\title[NSCs in late- and early-type hosts]
{
Masses and Scaling Relations for Nuclear Star Clusters, and their Coexistence with Central Black Holes
}
\author[I. Y. Georgiev, T. B\"oker, N. Leigh, N. L\"utzgendorf, N. Neumayer]
{Iskren Y. Georgiev$^1$\thanks{E-mail: georgiev@mpia.de ; iskren.y.g@gmail.com
}, Torsten B\"oker$^2$, Nathan Leigh$^{3,4}$, Nora L\"utzgendorf$^2$, \and \hspace{5cm} and Nadine Neumayer$^1$\\
$^{1}$Max-Planck Instiut f\"ur Astronomie, K\"onigstuhl 17, 69117 Heidelberg\\
$^{2}$Space Telescope Science Institute, 3700 San Martin Drive, Baltimore, MD 21218, USA\\
$^{3}$Department of Astrophysics, American Museum of Natural History, Central Park West and 79th Street, New York, NY 10024 \\
$^{4}$Department of Astronomy, University of Alberta, 116 St and 85 Ave, Edmonton, AB T6G 2R3, Canada\\
}

\begin{document}

\date{Accepted 2015 mm dd. Received 2015 mm dd}

\pagerange{\pageref{firstpage}--\pageref{lastpage}} \pubyear{2014}

\maketitle

\label{firstpage}

\begin{abstract}
Galactic nuclei typically host either a Nuclear Star Cluster (NSC, prevalent in galaxies with masses $\lesssim 10^{10}M_\odot$) or a Massive Black Hole (MBH, common in galaxies with masses $\gtrsim 10^{12}M_\odot$). In the intermediate mass range, some nuclei host both a NSC and a MBH. In this paper, we explore scaling relations between NSC mass (\nscmass) and host galaxy total stellar mass (${\cal M}_{\star,\rm gal}$) using a large sample of NSCs in late- and early-type galaxies, including a number of NSCs harboring a MBH. Such scaling relations reflect the underlying physical mechanisms driving the formation and~(co)evolution~of these central massive objects. We find $\sim\!1.5\sigma$ significant differences between~NSCs~in late- and early-type galaxies in the slopes and offsets of the relations \nscsize--\nscmass, \nscsize--${\cal M}_{\star,\rm gal}$ and \nscmass--${\cal M}_{\star,\rm gal}$, in the sense that $i)$ NSCs in late-types are more compact at fixed \nscmass\ and ${\cal M}_{\star,\rm gal}$; and $ii)$ the \nscmass--${\cal M}_{\star,\rm gal}$ relation is shallower for NSCs in late-types than in early-types, similar to the \bhmass--${\cal M}_{\star,\rm bulge}$ relation. 
We discuss these results in the context of the (possibly ongoing) evolution~of~NSCs,~de\-pen\-ding on host galaxy type. 
For NSCs with a MBH, we illustrate the possible influence of a MBH on its host NSC, by considering the ratio between the radius of the MBH sphere of influence and \nscsize. NSCs harbouring a sufficiently massive black hole are likely to exhibit surface brightness profile deviating from a typical King profile.
\end{abstract}

\begin{keywords}
galaxies: nuclei -- galaxies: star clusters -- galaxies: quasars: supermassive black holes
\end{keywords}

\section{Introduction}\label{Sect:Intro}

A growing body of observational evidence indicates that the nuclear regions of galaxies are often occupied by a nuclear star cluster (NSC) and/or a super massive black hole (SMBH), with NSCs being identified in more than $>60-70\%$ of early- \cite[e.g.][]{Durrell97,Carollo98,Geha02,Lotz04,Cote06,Turner12,denBrok14} and late-type galaxies \cite[e.g.][]{Boeker02,Boeker04,Balcells07a,Balcells07,Seth06,Georgiev09b,Georgiev&Boeker14,Carson15}. Driven by apparent similarities in the scaling relations of SMBHs and NSCs with host galaxy properties, \cite{Ferrarese06} introduced the term Central Massive Object\footnote{To clarify the terminology, we will use the tern CMO to describe the sum of NSC and MBH, if both are present.} (CMO), suggesting that the formation and evolution of both types of central mass concentration may be linked by similar physical processes. 

Indeed, the mass range of the two components of CMOs overlap, with SMBHs having \bhmass$\gtrsim\!10^6M_\odot$ \cite[e.g.][]{Gultekin09,Rusli13,McConnell&Ma13}, and NSC masses falling in the range $10^4 \lesssim$ \nscmass\, $\lesssim 10^8M_\odot$. Both \bhmass\, and \nscmass\, have repeatedly been found to correlate with a range of host galaxy properties including galaxy luminosity, mass, stellar velocity dispersion ($\sigma$), AGN activity etc. \cite[e.g.][and references therein]{Gultekin09,Seth08,Kormendy&Ho13}. These correlations followed the earlier discoveries that the mass of the SMBH scales with the host galaxy $B$-band bulge luminosity \citep{Kormendy&Richstone95}, dynamical mass \cite[e.g.][]{Magorrian98,Haering&Rix04}, stellar velocity dispersion \citep{Ferrarese&Merritt00,Gebhardt00} and central light concentration \cite[e.g.][]{Graham01}. 

Similar scaling relations are also found to hold between the mass of NSCs and their host galaxy bulge luminosity, mass \cite[e.g.][]{Ferrarese06,Wehner&Harris06} as well as morphological type \cite[e.g.][]{Rossa06,Erwin&Gadotti12}. The detailed shape of these relations possibly depends on host galaxy morphology, as suggested by \cite{Erwin&Gadotti12} who report a systematic difference in the NSC
mass fraction between early- and late-type hosts. It is, however, still hotly debated which is the fundamental physical mechanism setting these scaling relations \cite[e.g. gas accretion, cluster and/or galaxy mergers][]{Silk&Rees98,McLaughlin06,Li.Haiman.Low07,Leigh12,Antonini13}, or any combination of these \cite[see reviews by][]{Kormendy&Ho13,Cole&Debattista15}. Proper understanding of these issues is crucial for gaining insight into the formation and growth of CMOs, and, in turn, how a CMO might impact the evolution of the host galaxy. 

Perhaps one of the most intriguing observations is the coexistence of NSC and SMBH in galaxies with masses around ${\cal M}_{\rm gal}\simeq10^{10}M_\odot$  \citep{Filippenko&Ho03,Seth08,Seth10,Graham&Spitler09,Neumayer&Walcher12}, with the best-studied example being the center of the Milky Way \citep{Schoedel07,Ghez08,Gillessen09,Genzel10,Feldmeier14,Schoedel14b}. Throughout this paper, we will use the term ``coexisting'' whenever describing an NSC that contains a MBH. Finding coexisting NSCs and MBHs has triggered numerous studies to understand the nature of this co-existence and the processes involved in their formation, growth, mutual influence, and co-evolution \cite[e.g.][]{McLaughlin06,Li.Haiman.Low07,Nayakshin09,Bekki&Graham10}. 

For example, \cite{Neumayer&Walcher12} discuss the possibility that NSCs are susceptible to destruction by BHs when \bhmass\,/\,\nscmass\,$>> 1$, or when the MBH sphere of influence becomes comparable to the size of the NSC \citep{Merritt06}. Recent $N-$body simulations have demonstrated that the capture and accretion of stars migrating within the BH sphere of influence can significantly contribute to the mass growth of black holes as well as the central core density of the host galaxy \citep{Brockamp11,Brockamp14}. From theoretical arguments, the growth rate of MBHs is expected to increase with MBH mass and likely requires a seed BH with ${\cal M}>100M_\odot$, unless the BH host cluster is very dense \citep{Baumgardt04a,Baumgardt04b,Baumgardt05,Baumgardt06}. These simulations also show that a significant fraction of stars can escape from the cluster due to close encounters with the MBH \citep{Baumgardt04a,Baumgardt06}. 

The presence of a MBH can inhibit the onset of core-collapse in the NSC, and cause the NSC to expand, and ultimately to be disrupted \cite[e.g.][]{Merritt06,Merritt09,Tremaine95}. Depending on \bhmass\ and the cluster core density (concentration and core velocity dispersion), the impact of tidal stress forces from the MBH on the NSC will become significant at a radius comparable to that of the MBH sphere of influence, \smbhsize\ which scales linearly with \bhmass. Therefore, the effect of a MBH on the stellar orbits in the NSC is likely to be more pronounced in massive host galaxies (because more massive galaxies host more massive MBHs). This is true even in formation scenarios that involve the merging of systems: regardless of whether a NSC spirals into a nucleus that already contains a MBH, or whether a MBH falls into a nucleus occupied by a NSC, the structure and integrity of the NSC will be impacted if the MBH mass is a sufficiently high fraction of the bound NSC mass \cite[e.g.][ see also refs. in \S\,\ref{Sect:SMBH_NSC_coexistence}]{Antonini12,Antonini15,Antonini13} 

In massive globular clusters (GCs) and ultra compact dwarf galaxies (UCDs), the presence of MBHs is hotly debated. If confirmed, this would add support to the notion that some GCs may be the former nuclei of galaxies which lost significant amounts of mass in galaxy interactions/merging \cite[e.g.][and refs therein]{Luetzgendorf13,Mieske13,Seth14}. The (non-)presence of MBHs is therefore an extremely important factor for the study of the various types of compact stellar systems (NSCs, UCDs, and massive GCs), and possible evolutionary connections between them \cite[e.g.][]{Gregg09,Price09,Georgiev09,Georgiev09b,Georgiev12,Taylor10,Misgeld&Hilker11,Chiboucas11,Bruens11,Norris&Kannappan11,Foster11, Pfeffer&Baumgardt13,Puzia14,Georgiev&Boeker14,Frank14,Norris14,Seth14}.

Here, we explore scaling relations between the size/mass of NSCs and the stellar mass of their host galaxies, ${\cal M}_\star$, sorted by host morphology. In this context, it is reasonable to consider the total mass of the host galaxy (rather than just bulge mass). This is because the bulge mass of an early-type, elliptical galaxy is effectively equal to its total stellar mass, while in late-type galaxies, the bulge - if it exists at all - is negligible compared to the  disk component. Therefore, the main mass reservoir for NSC and/or SMBH formation in late-type disks would be ignored in studies that only consider the bulge mass of the host. While a few previous studies \citep{Carollo98,Erwin&Gadotti12} have taken the approach of considering the total host galaxy mass, our work significantly improves on the number of objects and the galaxy mass range, taking advantage of our recent catalogue of NSCs in disk galaxies \citep{Georgiev&Boeker14}. 

In \S\,\ref{Sect:Data Sample}, we describe the galaxy sample and the calculation of photometric masses for NSC and host galaxy. In \S\,\ref{Sect:Results_Analysis}, we present the analysis and comparison between late- and early-type galaxies using relations between NSC mass and size, \nscmass\,-\,\nscsize\ as well as between NSC properties and host galaxy stellar mass, \nscsize\,-\,${\cal M}_\star$ (\S\,\ref{Sect:NSC_mass_size_relation}), and \nscmass\,-\,${\cal M}_\star$ (\S\,\ref{Sect:NSC_gal_relations}). For nuclei with co-existing NSC and SMBH, we show in \S\,\ref{Sect:SMBH_NSC_coexistence} the corresponding relations for the combined CMO mass, ${\cal M}_{\rm BH+NSC}$\,-\,${\cal M}_\star$, and the ratio between the radius of the BH sphere of influence and the NSC effective radius, \smbhsize/\nscsize. The results are discussed in \S\,\ref{Sect:Discussion} and our conclusions are summarized in \S\,\ref{Sect:Conclussions}.

\section{Data samples and deriving NSC and galaxy photometric mass}\label{Sect:Data Sample}

\subsection{Morphological Sample Definitions}\label{Sect:Samples}
One of our main goals for this study is to check for differences in the CMO properties between early- and late-type galaxies, i.e. in nuclei of dynamically ``hot'' (bulge-dominated) and ``cold'', (disk-dominated) host galaxies. Such a morphology-based separation could indicate two different modes of evolution, e.g. active and inactive. Any observed differences between late- and early-type NSCs could therefore reflect the underlying environmental conditions for CMO formation \citep[e.g.][]{Leigh15}.

\begin{figure}
\includegraphics[width=.5\textwidth]{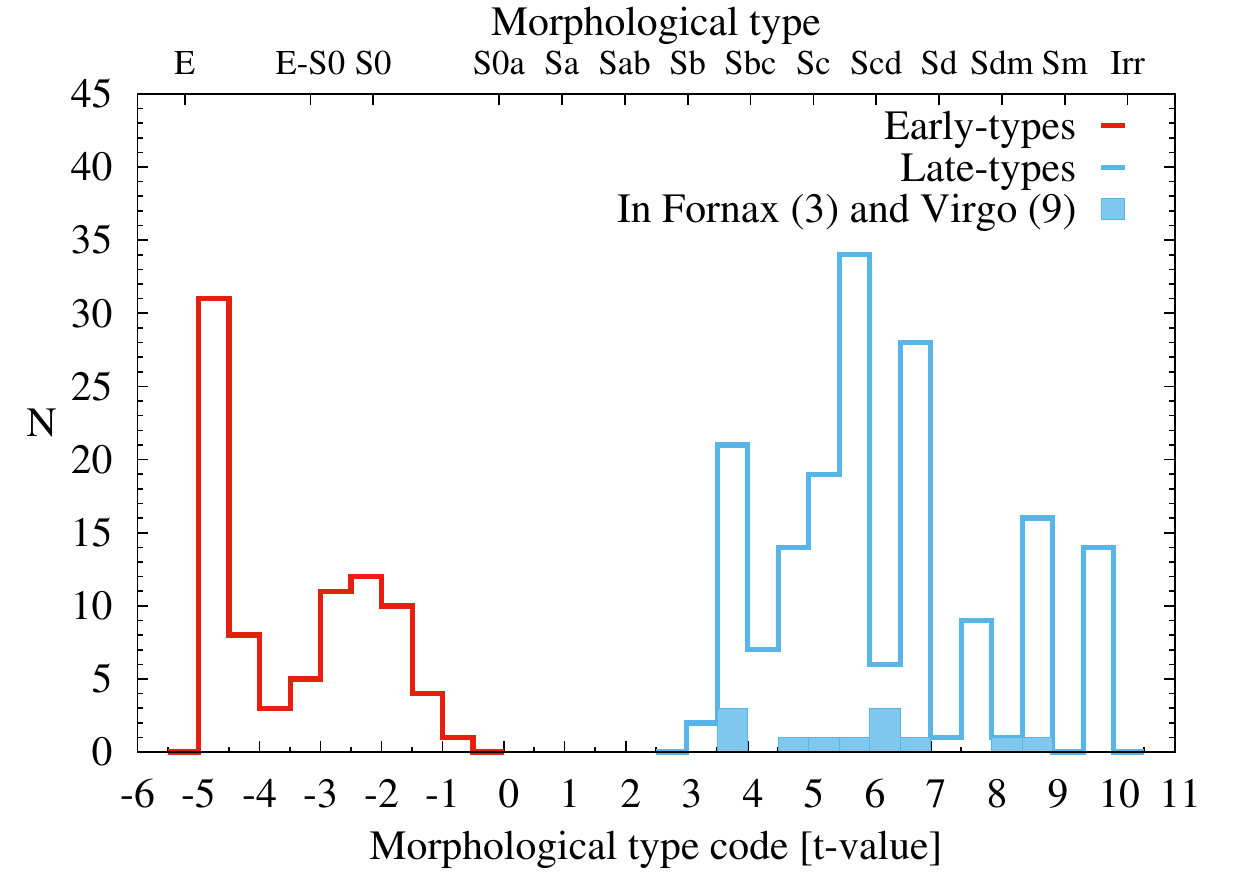}
\caption{Morphological distribution of the early- and late-type sub-samples. The t-type values are from HyperLEDA \protect\cite[based on][]{RC3}.
\label{fig:Morphology}
}
\end{figure}

To define our NSC sample, We use the $t\!-\!type$ galaxy morphological parameter defined by \cite{RC3}. The overall sample is comprised of NSCs in spheroid-dominated galaxies from \cite{Cote06} and \cite{Turner12}, as well as in disk-dominated galaxies from \cite{Georgiev&Boeker14} and \cite{Georgiev09b}. We divide this master sample into sub-samples of NSCs in early- and late-type hosts using the following criteria: the early-type sub-sample is comprised of all galaxies with $t\!<\!0$ (i.e. bulge dominated Es-S0s), while the late-type sub-samples contains all galaxies with $t\!>\!3$ (i.e. disk dominated, Sb and later). To demonstrate the clear separation of the two sub-samples, we show in Figure\,\ref{fig:Morphology} histograms of the t-type distribution within each subsample. This approach enables us to identify general trends and differences between the properties of NSCs in bulge and disk-dominated host galaxies, such as those reported by \cite{Erwin&Gadotti12} who find a change in the the mass ratio \nscmass/${\cal M}_{\star,\rm gal}$ occuring around $t\simeq3$ (their Fig.\,4), i.e. close to the morphological separation between our late- and early-type sub-samples.

We also note that our two sub-samples are dominated by galaxies in different environments. While virtually all NSC hosts in the early-type sample are located in a cluster environment (Virgo or Fornax), the late-type galaxies are found mainly in a lower density (group) environment, except for 12 galaxies ($<\!10\%$) that are members of Virgo or Fornax according to catalogues of \cite{Binggeli85} and \cite{Ferguson90}. We highlight these 12 objects with a solid histogram in Figure\,\ref{fig:Morphology}. 

Relevant \emph{only} for the NSC-MBH discussion in \S\,\ref{Sect:M_NSCBHvsM_gal} and \ref{Sect:BH.NSC_Sizeratio}, we also use data for galaxies hosting both a NSC and a MBH from \cite{Neumayer&Walcher12} and for SMBH host galaxies from \cite{McConnell&Ma13}. Both studies contain galaxies with a wide range of morphological types, and in order to separate these objects into early- and late-types, we adopt a galaxy morphology dividing line at $t\!=\!3$. 

\subsection{NSC photometry and mass}\label{Sect:Data_NSCs}

Deriving accurate photometric masses relies on properly accounting for foreground Galactic extinction to a given galaxy and precise knowledge of its distance. For this purpose, we retrieved 
the foreground Galactic extinction $E(B-V)$ and the (median value of the) distance modulus for all sample galaxies from NED\footnote{\href{http://ned.ipac.caltech.edu}{http://ned.ipac.caltech.edu}}. The NED extinction values are based on the \cite{Schlafly11} recalibration of the \cite{Schlegel98} extinction map, for which we calculate filter-specific values assuming the \cite{Fitzpatrick99} reddening law with $R_V\!=\!3.1$. The values for $E(B-V)$ and $m-M$ used for the computation of photometric stellar masses are listed in Table\,\ref{Table:GB_Gal_NSC}. 
We emphasize that possible NSC reddening due to host galaxy self-absorption is not accounted for. A correction for an $A_V\simeq0.4$\,mag  would increase the NSC mass by a factor of two (at fixed age and metallicity), which needs to be considered when estimating the systematic uncertainties (see also \S\,\ref{Sect:Data_NSCs} and discussion in \S\,\ref{Sect:NSC_gal_relations}).
The values for \nscmass\ and ${\cal M}_\star$ derived as described in the next sections are tabulated in Table\,\ref{Table:GB_Gal_NSC}\footnote{Full version of the table is available in the electronic version of the journal}.

The sample of NSCs in late-type galaxies in this work comes from the recently published catalogue of 228 NSCs in nearby ($\lesssim\!40$\,Mpc), moderately inclined spiral galaxies with $t\geq 3$ \citep{Georgiev&Boeker14}. These selection criteria ensure that the effects of any light contamination from the host galaxy disk and (pseudo-)bulge on the derived NSC properties are minimized. The catalogue contains luminosities calculated from the flux within the best fitting King model of a given concentration index. This provides the most accurate photometry in a nuclear environment because is less affected by nearby contaminating sources. 
 
The \cite{Bell03} mass-to-light ratio ($M/L$)-colour relations are available for the SDSS, 2MASS, and Johnson/Cousins magnitude systems. However, as discussed in \cite{Georgiev&Boeker14}, we prefer to work in the native WFPC2 magnitudes to avoid propagating uncertainties from transformations between the various photometric systems. For each NSC, we obtain the $M/L$-ratio using the NSCs magnitudes in \cite{Georgiev&Boeker14} and the \cite{BC03} SSP models for solar metallicity and a \cite{Kroupa01} IMF. As shown by spectroscopic studies of NSCs in late-type galaxies, the assumption of solar metallicity is a reasonable one for these objects \cite[e.g.][]{Rossa06,Walcher06,Seth06}. To calculate the luminosity weighted photometric mass of the NSCs, \nscmass, we used the available colour information in the various combinations of the most reliably calibrated WFPC2 filters ($F300W, F450W, F555W, F606W, F814W$). We obtain the SSP model $M/L$ by matching the NSC colours to the model colours. If more than one colour is available, we calculate the error weighted mean of the different $M/L$-values to obtain \nscmass, which helps to minimize systematic uncertainties \cite[cf][]{McGaugh&Schombert14}. 
For NSCs with photometry in only one band (i.e. without colour information), we used the sample median colours containing that filter to calculate the error weighted mean of the $M/L$ from the possible colour combinations containing that filter, e.g. for a NSC with only $F814W$ magnitude, we used the median $F300W\!-\!F814W,\ F450W\!-\!F814W$ and $F606W\!-\!F814W$ colours of the NSC sample to calculate the error weighted SSP model $M/L_{F814W}$. We checked, and expectedly, the calculated \nscmass\ using the sample median colours showed no systematic difference between those NSCs with mass calculated from measured colour(s).

Although these colours are representative for the entire sample of NSCs, we caution that there may still be a small bias in the NSC masses derived from different filters. We checked for this using NSCs observed in multiple filters, but did not find any systematic differences. We also note that NSCs with uncertainties larger than $>100\%$ in \nscmass\ (shown with gray symbols in subsequent figures) are excluded from the various fits\footnote{We found no significant differences in the best-fit parameters when including these NSCs in the fits, using weights that are inversely proportional to their uncertainty.}. 

\begin{figure}
\includegraphics[width=.5\textwidth, bb = 5 20 360 190]{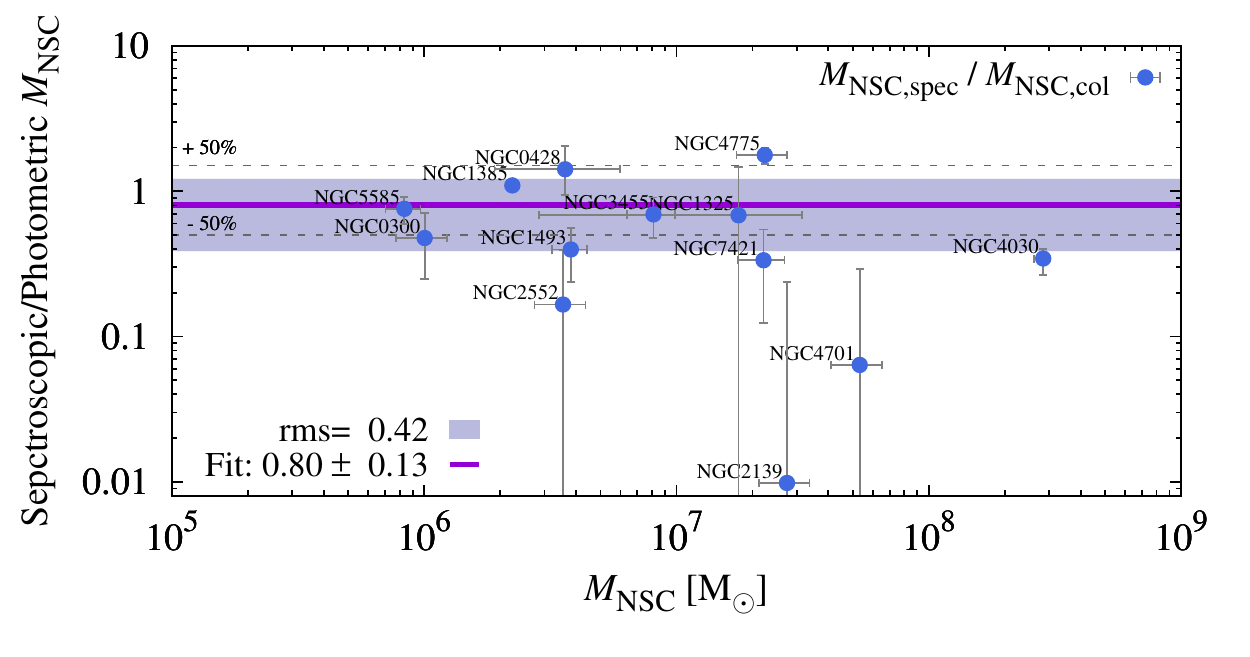}
\caption{Ratio between \nscmass\ measurements from spectroscopy \protect\cite[from][]{Rossa06} to those from photometric colours (this study). The error-weighted straight line fit (solid horizontal line) is plotted, together with the rms scatter of the data (shaded area). The dashed horizontal lines show the $\pm50\%$ range around a mass ratio of 1. The plotted errors are only from our study, as there are no error bars provided by \protect\cite{Rossa06}.
}\label{fig:Spec/Col_M_NSC}
\end{figure}

The formal errors of the measured colours are small due to the generally very high $S/N$ of the NSC, and thus introduce only negligible uncertainties in the resulting $M/L$-ratios. However, they  may still be affected by the possibility that a small mass fraction of the NSC ($\delta_{\cal M}\approx10\%$) is composed by a younger stellar population ($\Delta t\!>\!5$\,Gyr) which will outshine the more massive older stellar component. This will cause a bias towards bluer integrated colours, younger SSP ages, and a lower $M/L$ values, i.e. towards lower total \nscmass\ mass (by up to a factor of 5, see also discussion in \S\,\ref{Sect:NSC_mass_size_relation} and \ref{Sect:NSC_gal_relations}).  Our approach to derive \nscmass\ in late-type hosts is similar to that in \cite{Seth08} who find good agreement between photometric and dynamical mass estimates to within a factor of two, which is comparable to the mass uncertainties calculated here. 
Nevertheless, in Figure\,\ref{fig:Spec/Col_M_NSC} we illustrate how well our colour-based photometric NSC masses compare to those obtained from spectroscopic analysis (from stellar population fitting, \citeauthor{Rossa06}\,\citeyear{Rossa06}, or line widths, \citeauthor{Walcher06}\,\citeyear{Walcher06}). For those objects with both types of measurements, Figure\,\ref{fig:Spec/Col_M_NSC} plots their ratio as a function of NSC mass. The shaded area is the rms scatter of the data around the best fit ($\sigma\!=\!0.42$\,dex). For reference, dashed horizontal lines show the $\pm50\%$ range around a mass ratio of 1. The plot shows that within the uncertainties, both estimates are generally in good agreement. The photometric estimates appear to be higher by about 20\%, but the significance of this difference is only $<\!1.5\sigma$.
Nevertheless, a slight overestimation of the photometric mass could be expected if the assumed NSC metallicity is too high. For example, the $M/L$s would differ by about 20\% between solar and sub-solar metallicity in the \cite{BC03} SSP models. Thus, the systematic uncertainty in the photometric and spectroscopic mass estimates due to our choice of metallicity is much smaller than that caused by the degeneracy with stellar age, as discussed below.

The NSC photometry in early-type galaxies is collected from two galaxy cluster surveys conducted with with HST/ACS - the Virgo Cluster Survey \cite[ACSVCS,][]{Cote04} with photometry for 56 NSCs \citep{Cote06} and the Fornax Cluster Survey \cite[ACSFCS,][]{Jordan07a} with measurements of 31 NSCs \citep{Turner12}. The photometric mass of the NSCs of these samples are calculated in a similar way as for the late-types by using the $g_{F475W}-z_{F850LP}$ colour and $z$-band magnitude in the tables of \cite{Cote06} and \cite{Turner12}. 
We note that our approach in deriving \nscmass\ (from colours at fixed solar metallicity) differs from that adopted in those studies (\nscmass\ at fixed age of 5\,Gyr). According to the \cite{BC03} SSP models, at a fixed age of 5\,Gyr, the $M/L$ can vary by a factor of 2 with metallicity. For solar metallicity, the $M/L$ increases again by a factor of 2 between ages of 5 and 14\,Gyr. Therefore, either method carries an equal amount of uncertainty, roughly a factor of two. Our method therefore should yield \nscmass\ values that are consistent with other studies to within a factor of two.

\subsection{Properties of NSCs with massive BHs}\label{Sect:Data:NSC_SMBH}

For the subsample of NSCs with MBHs, we use the measurements of \cite{Neumayer&Walcher12} who provide upper limits for \bhmass\ based on velocity dispersions and dynamical mass modelling from VLT/UVES spectra.

Twelve NSCs (one late- and 11 early-type galaxies) in their sample are not in \cite{Georgiev&Boeker14}. For those NSCs, we use the luminosities obtained by \citeauthor{Neumayer&Walcher12} from a Multi-Gaussian Expansion fitting technique \cite[MGE, ][]{Emsellem94,Cappellari02}. The remaining seven NSCs in the \cite{Neumayer&Walcher12} sample are present in our HST sample, and we therefore use our photometry to calculate the NSCs masses, as explained in \S\,\ref{Sect:Data_NSCs}. To check for systematic differences between the two studies, we calculate the ratio between our and the \citeauthor{Neumayer&Walcher12} NSC sizes and masses. We find good agreement, with a mean ratios of $0.81\pm0.18$ for the NSC sizes, and $0.98\pm0.16$ for the NSC masses. The small apparent difference in derived sizes can likely be attributed to the different fitting techniques used, elliptical King~profiles~in \citeauthor{Georgiev&Boeker14} vs. MGE technique in \citeauthor{Neumayer&Walcher12}.

We also include in our sample the NSC and MBH masses of the Milky Way (MW) and Andromeda (M\,31). Their proximity makes these two nuclei the best-studied examples of systems with reliable mass measurements of both NSC and MBH. For the MW NSC, we use mass and size estimates from \cite{Schoedel14}, \nscmass\,$=(2.5\pm0.4)\times10^{7} M_\odot$, \nscsize\,$=4.2\pm0.4$\,pc. The mass of the MBH in the MW, \bhmass\,$=4.26\times10^6M_\odot$, is from \cite{Chatzopoulos14} based on stellar kinematics. For the total stellar mass of the MW,
we use the value derived by \cite{Licquia&Newman14}, ${\cal M}_\star=(6.08\pm1.14)\times10^{10}M_\odot$, which is based on an improved Bayesian statistical analysis accounting for uncertainties in literature measurements. The MW gas mass is ${\cal M}_{\rm MW gas}=1.25\times10^{10}M_\odot$ (about 17\% of its stellar mass), of which atomic Hydrogen constitutes ${\cal M}_{\rm HI}=8\times10^9M_\odot$, warm ionized medium ${\cal M}_{\rm H+}=2\times10^9M_\odot$, and molecular gas ${\cal M}_{\rm H2}=2\times10^9M_\odot$ \citep{Kalberla&Kerp09}. 

The M\,31 nucleus is a rather complex system \citep{Lauer93}. It is composed of a cluster that clearly stands out above the surrounding bulge within the central 10\,pc and is dominated by light from old stellar populations \citep{Kormendy&Bender99}. Its inner 1.8\,pc core features a bimodal component \citep{Lauer93,Lauer98,Lauer12}, which is interpreted as a projection of the Keplerian orbits of stars in a central eccentric disc around the MBH \citep{Tremaine95,Peiris&Tremaine03}. The mass of the MBH is \bhmass\,$=1.4\times10^8M_\odot$ \citep{Bender05} and that of the NSC \nscmass\,$=3.5\pm0.8\times10^7M_\odot$ \citep{Lauer98,Kormendy&Ho13}. We calculate the total stellar mass of M\,31 to be ${\cal M}_\star=7.88\pm4.23\times10^{10}M_\odot$, using $B,V,I$ photometry from HyperLEDA\footnote{\href{http://leda.univ-lyon1.fr}{http://leda.univ-lyon1.fr} \citep{Paturel03}} and using \cite{Bell03} $M/L$-colour relations (see also \S\,\ref{Sect:Data_Gal_Mags}). Our value for the M\,31 mass is consistent with other stellar population based estimates in the literature, e.g. 10-1$5\times10^{10}M_\odot$ \cite[][]{Tamm12}. 

\subsection{Sample of massive black holes}\label{Sect:Data_BHs}

Masses of 72 SMBHs and their host galaxies are taken from \cite{McConnell&Ma13}. They collect literature data from various sources of the most up to date \bhmass\ measurements. This sample is used in \S\,\ref{Sect:BH.NSC_Sizeratio} for calculating the SMBH sphere of influence radius, \smbhsize.

\subsection{Photometric stellar mass of NSC host galaxies}\label{Sect:Data_Gal_Mags}

We also calculate the total galaxy stellar mass (i.e. the sum of their bulge and disk components) for all NSC host galaxies in our sample. The total galaxy mass is an important quantity for the discussion of formation scenarios for both NSCs and SMBHs. This is especially true for late-type galaxies without prominent bulges, where material for the growth of either CMO must come predominantly from the disk.

Calculating galaxy photometric mass, ${\cal M}_{\star,\rm gal}$, from integrated colours in the optical is a challenging task, mainly due to the age-metallicity degeneracy and assumptions of galaxy star formation history used by synthetic models. However, it has been demonstrated that the $B\!-\!V$ colour (including for disk galaxies) offers a good representation of their stellar population \cite[e.g.][]{McGaugh&Schombert14}. We have therefore calculate ${\cal M}_{\star,\rm gal}$ using the empirically calibrated $M/L$-galaxy colour relations of \cite{Bell03}. These relations were obtained by comparing galaxy SEDs at optical and NIR wavelengths, and by using composite stellar evolutionary models for a range of metallicities and star formation histories.

For the majority of the galaxies in our samples, we collect photometry ($B, B\!-\!V, I$ magnitudes) from HyperLEDA, i.e. for the galaxies in \cite{Georgiev&Boeker14}, \cite{Turner12}, \cite{Neumayer&Walcher12} and \cite{McConnell&Ma13}. The only exception is the ACSVCS sample, for which we use the photometry derived from a dedicated isophotal analysis \citep{Ferrarese06a}. 

The \cite{McConnell&Ma13} catalogue provides host galaxy bulge stellar mass (either from spherical Jeans modelling of the bulge stellar dynamics, or from $M/L$-modeling based on galaxy colours, the latter approach being identical to ours). However, \cite{McConnell&Ma13} do not provide stellar mass estimates for those galaxies in their sample that contain a significant disk component, i.e. S0 and later types. For a consistent comparison to the NSC sample, we calculate the total (bulge+disk) stellar masses of their entire sample using magnitudes and colours obtained from HyperLEDA and the \cite{Bell03} $M/L$-color relations. To check the consistency of our results, we compared our galaxy masses to those of \cite{McConnell&Ma13} for the early-type galaxies in their sample, i.e. for cases where ${\cal M}_{\rm bulge}\simeq{\cal M}_\star$, and find a very good agreement (to within 10\%).
Uncertainties of the photometric masses have been calculated by propagation of the photometric uncertainties and the uncertainties associated to the coefficients of the $M/L$-color relation. To avoid over-crowding in the figures, galaxies with uncertainties larger than 100\% are shown with grey symbols.

\subsection{HI and X-ray gas masses}\label{Sect:HI_Xray_Masses}

A significant baryonic mass component in late-type galaxies is in the form of atomic HI gas. We therefore calculate the HI mass from the HyperLEDA 21-cm line magnitudes, $m21$, converted to flux ($F_{\rm HI}=10^{-0.4\times(17.40-m21)}$) using the relation between the ${\cal M}_{\rm HI}$ and $F_{\rm HI}$, i.e. ${\cal M}_{\rm HI}=2.36\times10^5\times D^2\times F_{\rm HI}$, where $D$ is the distance in Mpc, calculated from the same distance modulus in NED used for calculating galaxy mass from its luminosity. We note that we did not correct the HI mass for He fraction or molecular gas.

Early-type galaxies are known to contain a hot gas component, detected as an X-ray halo resulting from thermal Bremsstrahlung emission, which is known to trace well the total gravitating mass \cite[e.g.][]{Forman85,Fukazawa06}. Typically, the hot gas mass is no more than a few times $10^9M_\odot$ for a range of galaxy morphologies, environments, and luminosities ($L_K\simeq3-15\times10^{10}L_\odot$) \cite[e.g.][]{Bogdan13a,Bogdan13b,AndersonBD13}. This is only 6-7\% of the galaxy stellar mass \cite[e.g.][]{O'Sullivan03,Su&Irwin13} and is therefore not a significant component of the baryon mass budget. Nevertheless, for the early-type massive galaxies in the \cite{McConnell&Ma13} SMBH sample we collect the hot gas mass measured by \cite{Su&Irwin13}, based on Chandra and XMM data. Unfortunately, no X-ray measurements exist for most of our late-type sample, and we therefore do not list the hot gas mass fraction in Table\,\ref{Table:GB_Gal_NSC}. Given that the hot gas mass fraction is smaller in late-type galaxies than in ellipticals \cite{LiWangChen11}, and in any case accounts for only a small fraction of the total galaxy mass, this does not significantly affect our analysis or conclusions.

\section{Results and Analysis}\label{Sect:Results_Analysis}

Quantifying the relations between NSCs and their host galaxy, and possible dependence on galaxy morphology, bears important constraints for models of NSC formation and evolution, as well as for possible evolutionary connections to the various incarnations of compact stellar systems (e.g. massive GCs, UCDs). In addition, they provide insights into mechanisms that may transform galaxies from late- to early-type morphologies. With our large sample of NSCs in late-type hosts, we significantly increase the number of well studied nuclei in disk-dominated galaxies. This enables a more statistically meaningful comparison of \nscsize\ and \nscmass\ between early- and late-type galaxies, and extends the range of host galaxy masses to less massive systems.

\subsection{Relations between NSC size and NSC and host-galaxy masses}\label{Sect:NSC_mass_size_relation}

\begin{figure}
\includegraphics[width=.5\textwidth, bb=10 10 240 360]{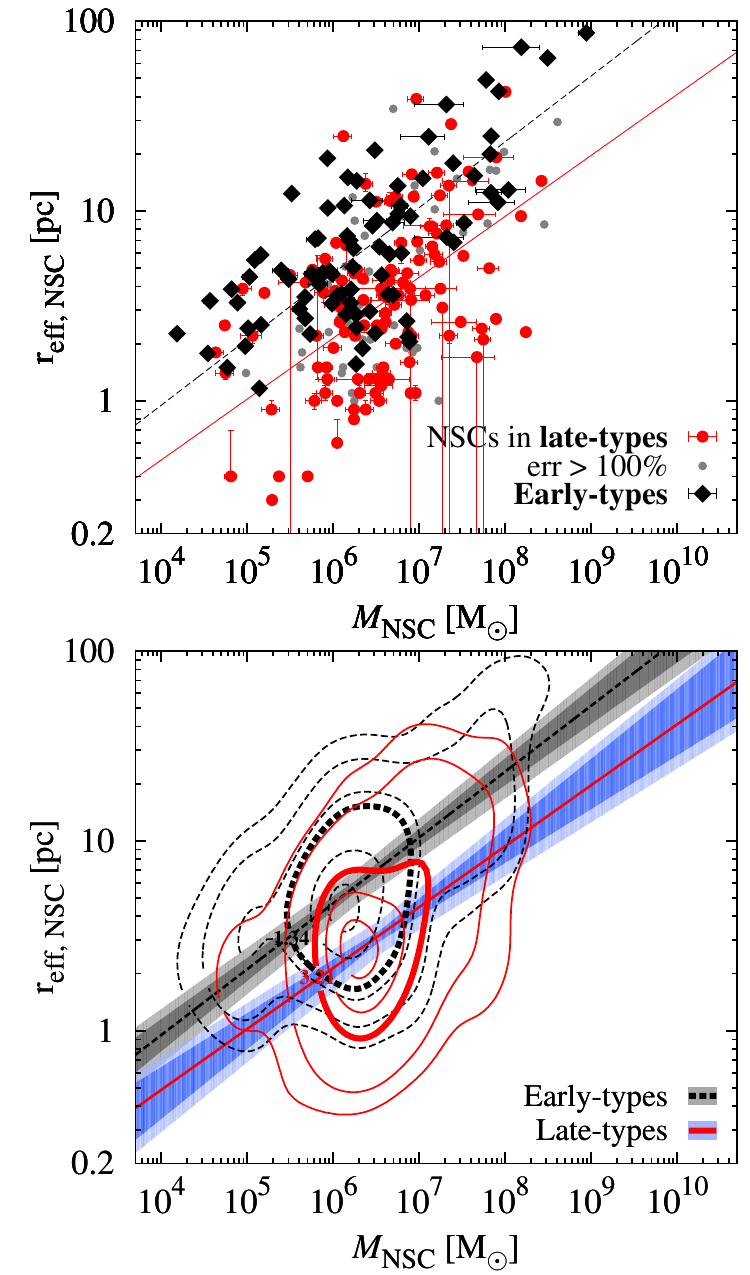}
\caption{{\bf Top:} Nuclear star cluster size - mass relation, \nscsize\ vs. -\nscmass\, for late- and early-type host galaxies. Small grey (light) symbols are for NSCs with uncertainties $>\!100\%$. {\bf Bottom:} a contour plot of the two-dimensional probability density distribution (2D\,PDF) for the two subsamples. The different symbols and line types are for the different samples, as indicated in the legend. Thick contour lines mark the $1\,\sigma$ of the 2D\,PDFs. The fit to the data is shown with lines, where the narrower darker (colour) shaded region indicates the uncertainties range of the fit slope and intercept. The wider and lighter (colour) shaded region is the $1\sigma$ dispersion of the data.  
}\label{fig:NSC_mass_size}
\end{figure}

\begin{figure}
\includegraphics[width=.5\textwidth, bb=10 10 240 360]{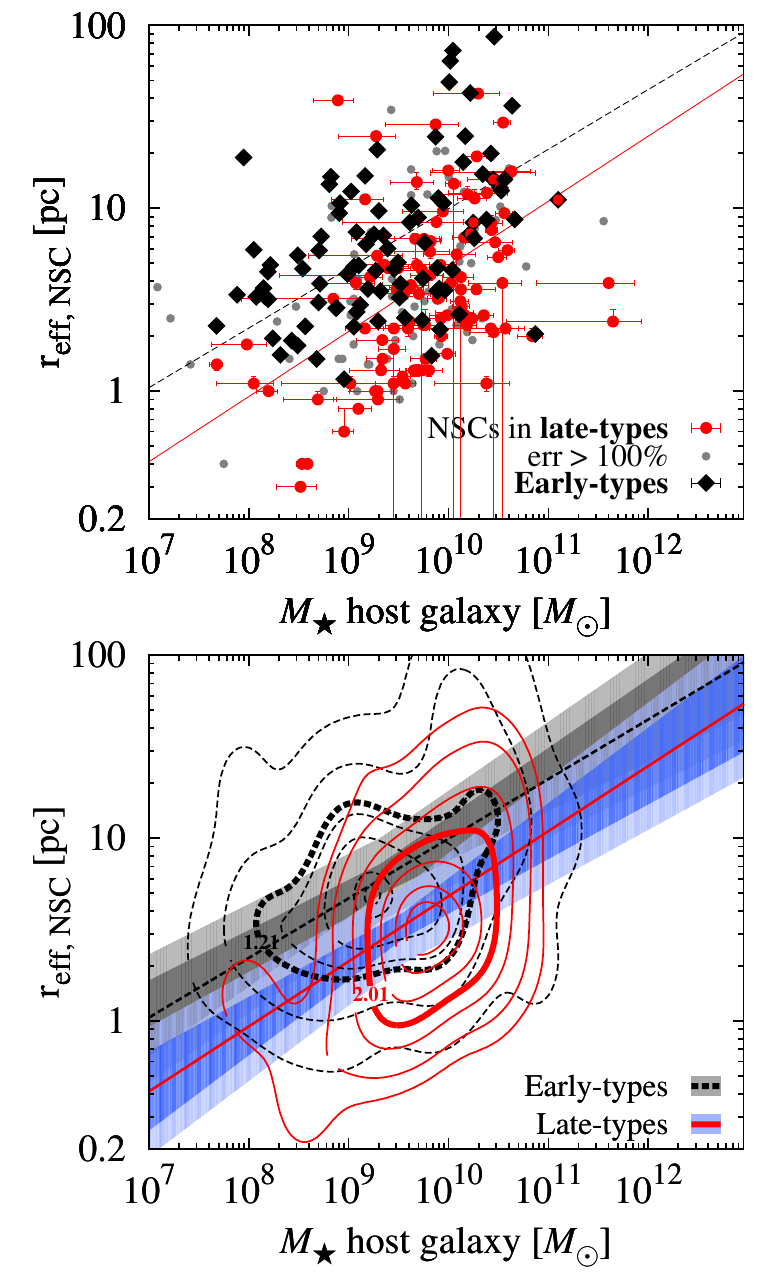}
\caption{{\bf Top:} Nuclear star cluster size versus total stellar mass of the host galaxy for NSCs in late- and early-type galaxies. {\bf Bottom:} A contour plot of the two-dimensional probability density distribution of the two subsamples. The symbol, line types and shaded areas are the same as in Fig.\,\ref{fig:NSC_mass_size}.
}\label{fig:Galaxy_NSCs_reff}
\end{figure}

The relations between the NSC effective radius and its mass (\nscsize\,-\,\nscmass) as well as the stellar mass of its host galaxy (\nscsize-${\cal M}_{\rm\star, gal}$) are shown in Figures\,\ref{fig:NSC_mass_size} and \ref{fig:Galaxy_NSCs_reff}, respectively. In both figures, late- and early-type host galaxies are shown with different symbol and line types, as indicated in the figure legend. We do not plot unresolved NSCs, i.e. those with only an upper limit to \nscsize\, \cite[see Fig.\,4 in][]{Georgiev&Boeker14}. In both figures, the bottom panel shows the two-dimensional probability density distribution function (2D-PDF). The uncertainty weighted 2D-PDFs are estimated within a running box of size 0.3\,dex within R\footnote{R is a free software environment for statistical computing. The R-project is an official part of the Free Software Foundation's GNU project (http://www.r-project.org/).}. The 2D-PDFs provide a first order quantification of the observed \nscsize-\nscmass\ distribution. The thick dashed and solid contour lines in Figures\,\ref{fig:NSC_mass_size} and \ref{fig:Galaxy_NSCs_reff} indicate the $1\sigma$ dispersion of the data.

To more robustly quantify any differences in the \nscsize-\nscmass\ and \nscsize--${\cal M}_{\rm\star, gal}$ relations between the two subsamples, we perform a maximum likelihood, linear (in log-log space) regression analysis by bootstrapping the data to account for the finite data sample and construct the posterior PDFs. Our fitting also accounts for the non-symmetric measurement uncertainties, which are treated as a combination of two Gaussians, i.e. a split normal distribution. The fitted linear regression is of the form:
\begin{equation}
\log_{10}($\nscsize$/c1)=\alpha\times\log_{10}($\nscmass$/c2)+\beta,
\end{equation}
where the normalization constants ($c1,\ c2$) and the best fit values for the slope ($\alpha$) and intercept ($\beta$) for the different subsamples are tabulated in Table\,\ref{Table:fits}. Description of the fitting technique and results for each relation are provided in the Appendix \S\,\ref{Sect:FitDetails}. 

\begin{deluxetable}{llllll}
\tabletypesize{\footnotesize}
%\rotate % For landscape mode
%\setlength{\tabcolsep}{0.15cm}
\tablecolumns{6}
\tablewidth{0pc} %%% <--- This is important!!! Otherwise it wont compile!!!!
\tablecaption{Parameters of the fitted relations for late- and \qquad\qquad\qquad\qquad\qquad\qquad\qquad\qquad\qquad\qquad\qquad\qquad\qquad\qquad\qquad\qquad
\\early-type NSC host galaxies.\qquad\qquad\qquad\qquad\qquad\qquad\qquad\qquad\qquad\qquad\qquad\qquad\qquad\qquad\qquad
\label{Table:fits}}
\tablehead{

\colhead{Host}&
\colhead{c1} &
\colhead{c2} &
\colhead{$\alpha$} &
\colhead{$\beta$} &
\colhead{$\sigma$} \\
\colhead{type} &
\colhead{} &
\colhead{} &
\colhead{} &
\colhead{} &
\colhead{} \\
\colhead{(1)} &
\colhead{(2)} &
\colhead{(3)} &
\colhead{(4)} &
\colhead{(5)} &
\colhead{(6)} \\
}
\startdata

						\multicolumn{6}{c}{}\\
\multicolumn{6}{l}{$\log($\nscsize$/c1)=\alpha\times\log($\nscmass$/c2)+\beta$} \\[.1cm]
{\bf Late}	&		3.31		&	3.60e6	&	$0.321^{+0.047}_{-0.038}$	&	$-0.011^{+0.014}_{-0.031}$	& 0.133	\\
{\bf Early}	&		6.27		&	1.95e6	&	$0.347^{+0.024}_{-0.024}$	&	$-0.024^{+0.022}_{-0.021}$	& 0.131	\\
						\multicolumn{6}{c}{}\\
\hline\\
\multicolumn{6}{l}{$\log($\nscsize$/c1)=\alpha\times\log({\cal M}_{\star,\rm gal}/c2)+\beta$} \\[.1cm]
{\bf Late}	&		3.44		&		5.61e9	&	$0.356^{+0.056}_{-0.057}$	&	$-0.012^{+0.026}_{-0.024}$	&	0.139	\\
{\bf Early}	&		6.11		&		2.09e9	&	$0.326^{+0.055}_{-0.051}$	&	$-0.011^{+0.015}_{-0.040}$	&	0.143	\\
						\multicolumn{6}{c}{}\\
\hline\\
%{\bf \nscmass\,--\,${\cal M}_{\star,\rm gal}$} & \multicolumn{5}{c}{} \\
\multicolumn{6}{l}{$\log($\nscmass$/c1)=\alpha\times\log({\cal M}_{\star,\rm gal}/c2)+\beta$} \\[.1cm]
{\bf Late}	&		2.78e6	&		3.94e9	&	$1.001^{+0.054}_{-0.067}$	&	$0.016^{+0.023}_{-0.061}$	&	0.127	\\
{\bf Early}	&		2.24e6	&		1.75e9	&	$1.363^{+0.129}_{-0.071}$	&	$0.010^{+0.047}_{-0.060}$	&	0.157	\\
						\multicolumn{6}{c}{}\\
\hline\\
%{\bf \nscmass\,--\,${\cal M}_{\star,\rm gal}$} & \multicolumn{5}{c}{} \\
\multicolumn{6}{l}{$\log($\nscmass$/c1)=\alpha\times\log({\cal M}_{\rm\star + HI, gal}/c2)+\beta$} \\[.1cm]
{\bf Late+HI}	&		2.87e6	&		1.17e10&	$1.867^{+0.158}_{-0.133}$	&	$0.041^{+0.041}_{-0.042}$	&	0.126 \\
						\multicolumn{6}{c}{}\\
\hline\\
%{\bf \nscmass\,--\,${\cal M}_{\star,\rm gal}$} & \multicolumn{5}{c}{} \\
\multicolumn{6}{l}{$\log({\cal M}_{\rm NSC + MBH}/c1)=\alpha\times\log({\cal M}_{\rm\star, gal}/c2)+\beta$} \\[.1cm]
{\bf MBH+NSC}	&		5.03e7&		2.76e10&	$1.491^{+0.149}_{-0.097}$	&	$-0.019^{+0.111}_{-0.054}$	&	0.233 \\
						\multicolumn{6}{c}{}\\
%$M_{\rm NSC}$	&	\multicolumn{4}{c}{}	\\
%early-types w/o HI		&	$1.84\pm0.51$	&	$9.99\pm0.71$	&		\nodata		&		\nodata		\\
%late-types w/o HI		&	$1.84\pm0.47$	&	$9.04\pm0.68$	&		\nodata		&		\nodata		\\
%late-types with HI		&	$2.71\pm0.36$	&	$8.87\pm0.30$	&		\nodata		&		\nodata		\\
%all types w/o HI		&	$1.94\pm0.52$	&	$10.04\pm0.75$	&		\nodata		&		\nodata		\\
%						\multicolumn{5}{c}{}\\
%$M_{\rm NSC + BH}$	&	\multicolumn{4}{c}{}	\\
%all types w/o HI		&	$1.32\pm0.11$	&	$8.50\pm0.11$	&		\nodata		&		\nodata		\\
%all types with HI		&	$1.08\pm0.47$	&	$7.95\pm0.29$	&		\nodata		&		\nodata		\\
%						\multicolumn{5}{c}{}\\
%$M_{SMBH}$		&	\multicolumn{4}{c}{}	\\
%all types w/o HI		&	$1.37\pm0.23$	&	$8.19\pm0.10$	&		\nodata		&		\nodata		\\
%all types with HI		&	$0.84\pm0.23$	&	$7.94\pm0.15$	&		\nodata		&		\nodata		\\
%early-types w/o HI		&	$0.76\pm0.18$	&	$8.56\pm0.09$	&		\nodata		&		\nodata		\\
%late-types with HI		&	$0.75\pm0.37$	&	$6.58\pm0.29$	&		\nodata		&		\nodata		\\
\enddata

\vspace{-.5cm}
\tablecomments{
The fitted scaling relations are of the form $\log_{10}(y/c1)=\alpha*\log_{10}(x/c2)+\beta$, where in column (1) is the NSC host morphological type, columns (2) and (3) are the normalization constants obtained from the 2D PDFs (see \S\,\ref{Sect:NSC_mass_size_relation}), in columns (4) and (5) are the slope and intercept and in (6) is the fit rms dispersion of the data, $\sigma$.
}

\end{deluxetable}

The values of the normalization constants are the highest probability density value of the center peaks in the contour plot in Fig.\,\ref{fig:NSC_mass_size}. We find that these normalization constants are important for minimizing the correlation between the slope and intercept, which provides a more realistic uncertainty estimate of the fits. The posterior probability density distributions of the slope and intercept for each fitted relation are shown in Figure\,\ref{fig:A.fits} in \S\,\ref{Sect:FitDetails}. We find that
$\log$\nscsize\ scales with $\log$\nscmass\ with a slope of $\alpha=0.321^{+0.047}_{-0.038}$ for late-types and $0.347^{+0.024}_{-0.024}$ for early-types. The $\log$\nscsize\ is also observed to scale with host galaxy stellar mass with a slope of $\alpha=0.356^{+0.056}_{-0.057}$ for late-types and $0.326^{+0.055}_{-0.051}$ for early-types.

The comparison between the 2D-PDFs of NSCs in the late- and early-type subsamples suggests that the \nscsize--\nscmass\ distributions are consistent with each other to within $1\sigma$ of the dispersion of the data (cf. the solid $1\sigma$ contour lines in Fig.\,\ref{fig:NSC_mass_size}, bottom). Within the uncertainties, the relations for late- and early-type hosts also have very similar slopes, however, the zeropoint of the fitted relations differ beyond their $1\sigma$ dispersion (cf. the broader shaded region around the highest density peaks in Fig.\,\ref{fig:NSC_mass_size}). This suggests that at fixed cluster mass, NSCs in late-type hosts are smaller by about a factor of 2 than their counterparts in early-type hosts. 

A similar difference in \nscsize\ between late- and early-type galaxies is also observed as a function of galaxy stellar mass, (\nscsize-${\cal M}_{\rm\star, gal}$), shown in Figure\,\ref{fig:Galaxy_NSCs_reff}. The \nscsize\ increases with ${\cal M}_{\rm\star, gal}$ with an identical slope for both samples, however, at fixed ${\cal M}_{\rm\star, gal}$, NSCs in late-type hosts are more compact by about a factor of 2. In this case, however, the statistical significance that both distributions differ is less than $1\sigma$, for the offset, slope and the 2D density distributions of the data (cf fits' shaded regions and solid density contours in Fig.\,\ref{fig:Galaxy_NSCs_reff}). These differences are discussed in \S\,\ref{Sect:Discussion:differences}.

\subsection{NSC mass -- host galaxy stellar mass relation}\label{Sect:NSC_gal_relations}

\begin{figure*}
\includegraphics[width=1.05\textwidth, bb=20 20 820 230]{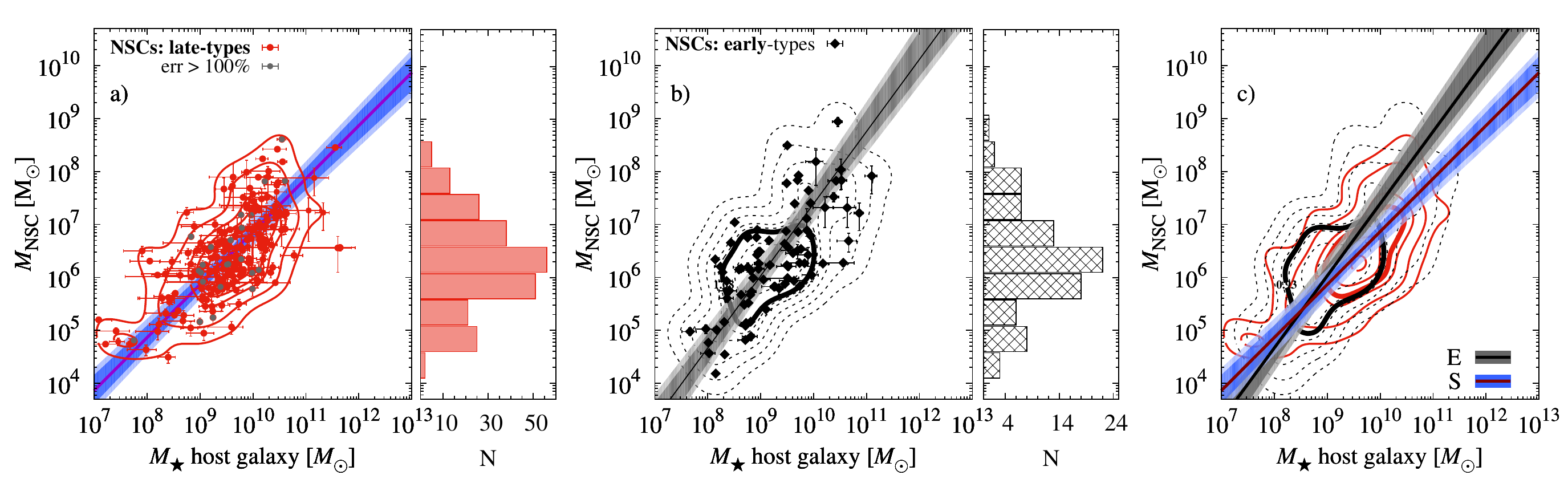}
\caption{Relation between nuclear star cluster mass, \nscmass, and host galaxy stellar mass, ${\cal M}_{\rm \star, gal}$. {\bf Panels a) and b)} show separately the relations for late- and early-type galaxies. The histograms on the y2-axes show the NSCs mass distributions for the different samples. In {\bf panel c)} we compare the fitted relations from panels a) and b) and their 2D PDF distribution (E for early- and S for late-types). Thick contour lines indicate the $1\sigma$ of the data PDF. Symbols, line types and shaded areas are the same as in Fig.\,\ref{fig:NSC_mass_size}. 
}\label{fig:Gal_NSCs_mass}
\end{figure*}

In Figure\,\ref{fig:Gal_NSCs_mass}, we explore the relation between the NSC mass and host galaxy stellar mass, again separately for late- (Fig.\,\ref{fig:Gal_NSCs_mass}\,a) and early-type galaxies (Fig.\,\ref{fig:Gal_NSCs_mass}\,b). We fit the two subsamples with the same technique as described in \S\,\ref{Sect:NSC_mass_size_relation}.
The best-fit relations are shown with solid lines, while the shaded regions represent the uncertainties of the fit coefficients (the narrower, darker region) and the $1\sigma$ dispersion of the data (the broader, lighter region). The direct comparison between the relations for late- and early-type NSC host galaxies in Figure\,\ref{fig:Gal_NSCs_mass}\,c shows that within $1\sigma$, their 2D-PDFs (thick contour lines) are indistinguishable from each other. On the other hand, the comparison also shows that the fitted slopes (solid and dashed lines in Fig.\,\ref{fig:Gal_NSCs_mass}\,c) are different between the two sub-samples beyond the $1\sigma$ level (i.e. the darker shaded region in Fig.\,\ref{fig:Gal_NSCs_mass}\,c do not overlap). This implies that at higher galaxy mass, early-types have more massive NSCs than late-types. A similar difference as a function of galaxy morphology has been also reported earlier \citep{Rossa06,Seth08,Erwin&Gadotti12}.

The values of the best-fit coefficients are summarized in Table\,\ref{Table:fits}. Our result that the mass of the NSC scales with host galaxy stellar mass with a slope near unity for late-types, $\alpha=1.001^{+0.054}_{-0.067}$, is in good agreement with the literature, e.g. \cite{Erwin&Gadotti12}, who derive a slope of $\alpha=0.90\pm0.21$ between \nscmass\ and total (bulge plus disk) stellar galaxy mass in a smaller sample of massive late-type spirals. 

We note at this point that the slope of the relation for the late-types is similar to the slope defined by the MBH mass and host spheroid mass, \bhmass--${\cal M}_{\rm \star, sph}$, which is $\alpha=1.05$ \citep{McConnell&Ma13}. However, the \bhmass--${\cal M}_{\rm \star, sph}$ has a zeropoint of 8.46, which is 0.6\,dex higher compared to the $7.86\pm0.1$ for our late-types relation. In \S\,\ref{Sect:Discussion:differences} we discuss first whether the differences between the relations for late- and early-types are due to measurement biases or evolutionary differences, and then discuss the implications for the \cmomass--${\cal M}_{\star\rm, gal}$ relation.

\begin{figure*}
\includegraphics[width=1.05\textwidth, bb=20 20 770 230]{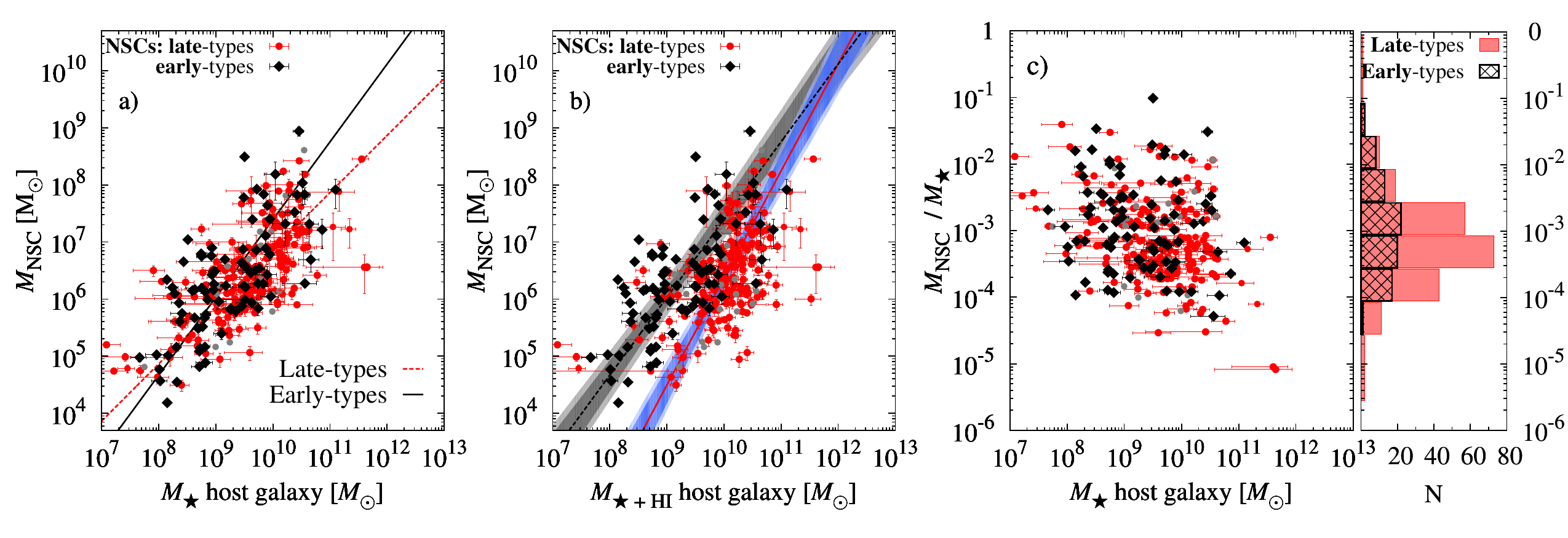}
\caption{{\bf Panel a)} \nscmass--${\cal M}_{\rm\star, gal}$ relation and in {\bf panel b)} with added HI mass, ${\cal M}_{\rm\star + HI, gal}$. NSCs in late- and early-type galaxies are shown with different symbol types and colours as indicated in the figure legend. For reference, with lines in panel a) we show the same fits as in Fig.\,\ref{fig:Gal_NSCs_mass}, whereas in panel b) are the fits including the HI mass. The fit values for the different samples are given in Table\,\ref{Table:fits}. {\bf Panel c)} shows the mass of NSC compared to host galaxy stellar mass. Symbols, line types and shaded areas are the same as in Fig.\,\ref{fig:NSC_mass_size}. 
}\label{fig:Gal_NSCs_mass_all}
\end{figure*}

It is perhaps equally interesting to see how the \nscmass--${\cal M}_{\rm\star, gal}$ relation changes when including the HI mass to the total host galaxy mass. Naturally, the effect on the \nscmass--${\cal M}_{\rm\star, gal}$ relation will be larger for gas-rich late-type galaxies. To gauge the magnitude of this effect, we show in Figure\,\ref{fig:Gal_NSCs_mass_all}\,a the NSC mass distribution for both early- and late-types against host galaxy stellar mass, ${\cal M}_{\rm\star, gal}$, and in Figure\,\ref{fig:Gal_NSCs_mass_all}\,b against the total galaxy mass, ${\cal M}_{\star + \rm HI}$. To guide the eye, we overplot again the best-fit relations from Fig.\,\ref{fig:Gal_NSCs_mass} in Figure\,\ref{fig:Gal_NSCs_mass_all}\,a. It is evident from Figure\,\ref{fig:Gal_NSCs_mass_all}\,b that when the HI mass is included, the relation for late-types steepens significantly. This is mostly because 
the low mass, late-type, galaxies have the highest HI mass fraction, and thus move noticeably to the right (i.e. toward higher total mass) in Figure\,\ref{fig:Gal_NSCs_mass_all}\,b, causing the relation to steepens. Due to the purely illustrative purposes of this comparison, we did not attempt to include He or molecular mass corrections. Those will only further strengthen the differences. As mentioned in \S\,\ref{Sect:HI_Xray_Masses}, ignoring the small fraction ($<\!5\%$) of the total galaxy mass contained in X-ray emitting hot gas mass should not significantly affect our results.

In Figure\,\ref{fig:Gal_NSCs_mass_all}\,c we plot \nscmass/${\cal M}_{\star \rm gal}$, i.e. the fraction of galaxy mass contained in the NSC. Overall, both for late- and early-type hosts, the mass of the NSC is about 0.1\% of the galaxy stellar mass (\nscmass/${\cal M}_{\star \rm gal}\rm\simeq10^{-3}$, cf the histogram in Fig.\,\ref{fig:Gal_NSCs_mass_all}\,c with a dispersion of about a factor of three. The slope of the \nscmass-${\cal M}_{\rm gal}$ relation adds to the broadening of the histogram projections. The observed NSC mass fractions in this late-type host galaxy sample are consistent with the values (0.1-0.2\%) reported by earlier studies \cite[e.g.][]{Rossa06,Seth08,Graham&Spitler09,Erwin&Gadotti12,Kormendy&Ho13}. We note that \cite{Erwin&Gadotti12} reported \nscmass/${\cal M}_{\star\rm gal}\sim0.2\%$ for Hubble types earlier than Sbc (consistent with our results), but $\sim0.03\%$ for later Hubble types, which is lower than the peak of the distribution in this study. However, partly due to the slope of the \nscmass\,-${\cal M}_{\star\rm gal}$ relation the \nscmass/${\cal M}_{\star\rm gal}$ distribution shows a large dispersion in Fig.\,\ref{fig:Gal_NSCs_mass_all}\,c with a $1\sigma$ range between $6\times10^{-6}$ (0.006\%) and $3\times10^{-3}$ (0.1\%).

\subsection{Relations for coexisting NSCs and MBHs}\label{Sect:SMBH_NSC_coexistence}

The identification and study of systems in which NSC and MBH coexist is important in order to make progress on a number of open questions. For example, it is not clear whether this coexistence is possible only in the nuclei of intermediate mass galaxies (few\,$\times10^{10}M_\odot$), or what the physical reason is for the dominance of one or the other at low and high galaxy mass. Understanding whether there is a common scaling relation for NSC and MBHs with host ${\cal M}_\star$ promises to shed light on the processes that govern their growth, i.e. the processes funnelling matter (gas, stars, star clusters) towards the deepest point of the host galaxy potential \cite[e.g.][]{Li.Haiman.Low07,Mayer10,Hartmann11,Antonini12,Antonini15}, and the feedback processes affecting the growth by either NSC or MBH.

Observationally, however, it is extremely challenging to populate the respective scaling relations, mostly because in the absence of accretion activity, the dynamical effects of a low-mass MBH on the surrounding NSC are below the detection threshold of current instruments. On the high-mass end, one could ask why NSC are not observed around SMBHs with \bhmass$>10^8M_\odot$? To address these questions, we first look at the combined mass of the NSC and MBH in coexisting systems \citep{Neumayer&Walcher12}, with an eye on the impact of a MBH that is massive enough to affect more than 50\% of the NSC.

\subsubsection{${\cal M}_{\rm NSC+BH}$ - host galaxy stellar mass relation}\label{Sect:M_NSCBHvsM_gal}

\begin{figure}
\includegraphics[width=0.5\textwidth, bb=20 20 370 250]{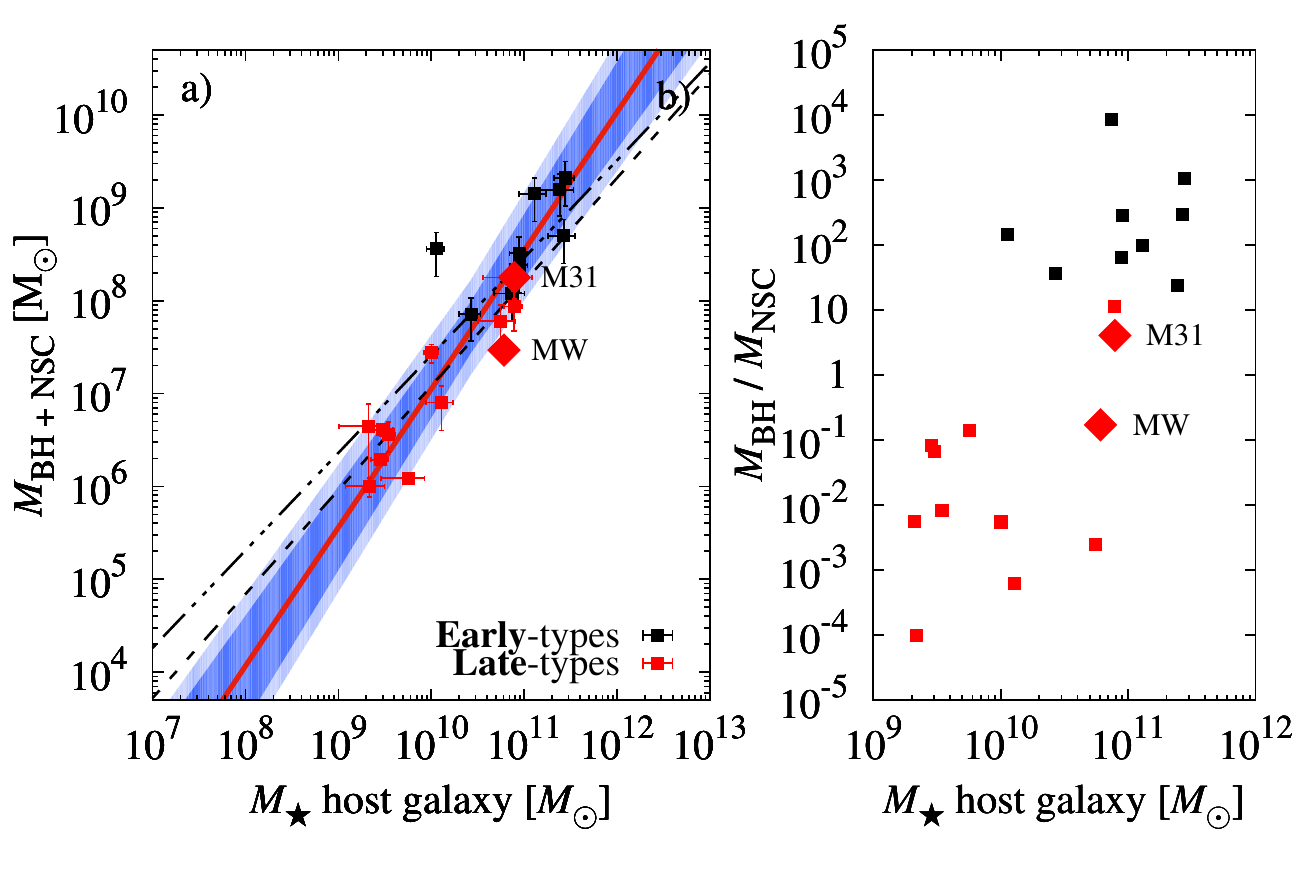}
\caption{Sum of the mass of NSC with MBH ({\bf panel a}) and their mass ratio  ({\bf panel b}) against host galaxy stellar mass (data from \protect\cite{Neumayer&Walcher12}). The different symbol types indicate late- and early-type galaxies, as indicated in the legend. Labeled also are the Milky Way and M\,31. The fit through the data in panel\,a) is shown with solid line and the shaded region indicates the uncertainty of the fit values and the rms of the data. For reference, dashed line is the \protect\cite{McConnell&Ma13} \bhmass-${\cal M}_{\rm bulge}$ relation and dash-dotted line is the \nscmass-${\cal M}_\star$ relation for late-type galaxies obtained in \S\,\ref{Sect:NSC_gal_relations} (discussion in \S\,\ref{Sect:SMBH_NSC_coexistence}\,and\,\ref{Sect:Discussion:CMO}).
}\label{fig:MBH_NSC}
\end{figure}

In Figure\,\ref{fig:MBH_NSC}, we show the combined mass of the 
CMO (i.e. \nscmass\ + \bhmass) against host galaxy stellar mass for late- and early-type galaxies, plotted with light and dark symbols, respectively. The values for \nscmass\ are calculated from luminosities in \cite{Georgiev&Boeker14} and \cite{Neumayer&Walcher12} as described in \S\,\ref{Sect:Data_NSCs} and \ref{Sect:Data:NSC_SMBH}.

The maximum likelihood, bootstrapped, non-symmetric error weighted fit is shown with a solid line in Figure\,\ref{fig:MBH_NSC}\,a. As before, the shaded regions indicate the uncertainty of the fit and the $1\sigma$ dispersion of the data. The fit values are listed in Table\,\ref{Table:fits}. For comparison, we overplot the \bhmass-$M_{\rm bulge}$ relation from \citet{McConnell&Ma13} with a dashed line, and with a dash-dotted line the \nscmass-${\cal M}_{\star\rm gal}$ relation for late-type galaxies obtained in \S\,\ref{Sect:NSC_gal_relations} (cf Fig.\,\ref{fig:Gal_NSCs_mass}).  We find that the sum of the NSC and MBH masses also defines a relation with host galaxy stellar mass, with a slope of $\alpha=1.491^{+0.149}_{-0.097}$. This slope is similar to that of the early-type \nscmass--${\cal M}_{\rm\star gal}$ relation ($\alpha=1.363^{+0.129}_{-0.071}$), but significantly steeper than the one for late-type hosts. We note that the ${\cal M}_{\rm NSC+MBH}-{\cal M}_{\star,\rm gal}$ is steeper than the \bhmass-${\cal M}_{\rm bulge}$ relation \cite[e.g. $\alpha=1.05\pm0.11, 1.12\pm0.06$, respectively][]{McConnell&Ma13,Haering&Rix04}. In their sample of massive late-type spirals, \cite{Erwin&Gadotti12} find a slope of $\alpha=1.27\pm0.26$ for the \bhmass-$M_{\star\rm bulge}$ relation, but unfortunately, they do not provide a fit against total galaxy mass. 

The fact that the \nscmass-${\cal M}_{\star\rm gal}$ and \bhmass-$M_{\rm bulge}$ relations have similar zeropoint and slope (cf. dash-dotted and dashed lines in Fig.\,\ref{fig:MBH_NSC}\,a) is perhaps not surprising, given that the bulge mass of early-type galaxies is a good approximation for the total galaxy stellar mass. In \S\,\ref{Sect:Discussion:CMO}, we further discuss these relations in the context of the coexistence of 
NSC and MBH and the transition from one to the other.

In Figure\,\ref{fig:MBH_NSC}\,b, 
we plot the mass ratio between the NSC and MBH, \bhmass/\nscmass, against host galaxy stellar mass. The plot shows that at total stellar host masses around $5\times10^{10}M_\odot$, the BH mass begins to dominate over the NSC mass, while for lower galaxy masses, the NSC outweighs the MBH. Mass ratios of $\ll\!1$ in late-type galaxies were first pointed by \cite{Seth08}. Subsequently, \cite{Graham&Spitler09} also included data for coexisting NSCs and MBHs in early-types, but they considered the fractional mass ratio \bhmass/(\nscmass+\bhmass) against host spheroid mass. \cite{Neumayer&Walcher12}, whose data we use here, plot directly \nscmass\,vs.\bhmass. From the lack of correlation between the two, they conclude that NSCs and BHs do not correlate as strongly with each other as they do with their host galaxy. 

\subsubsection{MBH to NSC size ratio}\label{Sect:BH.NSC_Sizeratio}

There has been a large body of analytical and numerical work to understand the effect on the formation and evolution of a NSC due to the presence of a MBH \cite[e.g.][and many others]{Tremaine95,Milosavljevic&Merritt01,Peiris&Tremaine03,Merritt06,Merritt09,Baumgardt05,Baumgardt06,Matsubayashi07,Bekki&Graham10,Brockamp11,Antonini13,Lupi14,Mastrobuono-Battisti14,Antonini12,Antonini15}. In this section, we attempt to explore this topic from the observational perspective by using the MBH sphere of influence (\smbhsize) and the effective (half-mass) radius of the NSC (\nscsize). These are observables that can be linked to theoretical expectations and provide observational information/expectation as to whether a significant fraction of the NSC stars/mass is influenced by the MBH. For an isotropic, virialized stellar cluster, the size ratio between NSC and MBH is effectively equivalent to their mass ratio, because in such an idealized system, \smbhsize\,$=\!G$\bhmass$/\sigma^2$ and \nscsize\,$=\!G$\nscmass /$\sigma^2$, and hence \smbhsize/\nscsize\,$\equiv$\,\bhmass/\nscmass . When \smbhsize/\nscsize$=1$, all stars within \nscsize\ are strongly bound to the MBH and have mostly Keplerian orbits. Thus, beyond the \nscsize, the cluster will have profile represented by a King model. Therefore, at this limit, the inner 50\% of the NSC potential (i.e. the stellar orbits) are dominated by the MBH, and the outer 50\% are dominated by the NSC potential/mass distribution. It follows that, for \smbhsize/\nscsize$>>1$, the ``classic'' NSC SB profile should no longer exist, and the NSC potential should be entirely dominated and shaped by the MBH. In other words, it is reasonable to expect that when \smbhsize/\nscsize$>>1$, the NSC integrity may be compromised, to the point that the very definition of an NSC may change, both theoretically and observationally.

\begin{figure*}
\includegraphics[width=1.\textwidth, bb= 10 20 700 250]{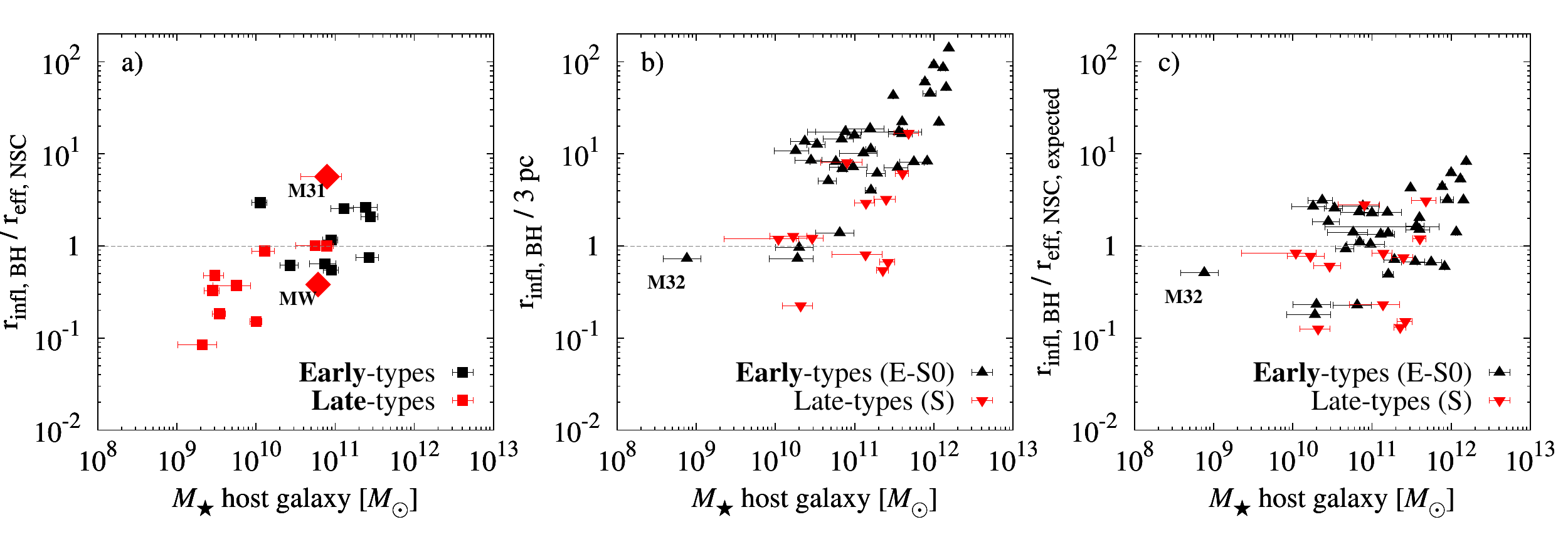}
\caption{The ratio between the MBH sphere of influence radius, \smbhsize, and the NSC effective radius, \nscsize, against host galaxy stellar mass, ${\cal M}_{\star\rm gal}$. {\bf Left:} size ratios from directly measured quantities \citep{Neumayer&Walcher12}. {\bf Middle:} shows the ratio \smbhsize~/~3\,pc for galaxies with SMBHs from \protect\cite{McConnell&Ma13} (details in \S\,\ref{Sect:Data_BHs} and \ref{Sect:SMBH_NSC_coexistence}). {\bf Right:} ratio of the \smbhsize\ to a predicted \nscsize\ according to the \nscsize--${\cal M}_{\rm\star, gal}$ empirical relation derived in \S\,\ref{Sect:NSC_mass_size_relation}, Figure\,\ref{fig:Galaxy_NSCs_reff}. Dashed, grey line in all panels indicates a ratio of unity. Late- and early-type galaxies are shown with different symbols as indicated in the legend. For reference, the Milky Way, M\,31 and M\,32 are indicated with labels.
}\label{fig:Gal_mass_rBH_toNSC}
\end{figure*}

In what follows, we derive these two characteristic sizes for NSCs and MBHs, first for nuclei in which both are known to coexist. We calculate the radius of the BH sphere of influence as \smbhsize$=G$\bhmass$/\sigma^2$, where \bhmass\ is from \cite{McConnell&Ma13} and \cite{Neumayer&Walcher12} and $\sigma$ is the central velocity dispersion taken from HyperLEDA for the \citeauthor{McConnell&Ma13} sample, or as measured by \citeauthor{Neumayer&Walcher12} for their NSC-MBH sample. We note that the $\sigma$ values from HyperLEDA may be biased toward higher values due to the often limited spatial resolution in measuring $\sigma$. Our calculated \smbhsize\ values for the \citeauthor{McConnell&Ma13} sample may be affected by this bias.

In Figure\,\ref{fig:Gal_mass_rBH_toNSC}\,a we plot the ratio \smbhsize /\nscsize\ against host galaxy stellar mass. For reference, the data points for the Milky Way (MW) and M\,31 are labelled, and the unity ratio is indicated with a dashed horizontal line. As expected, the MW has a size ratio below one, while M\,31 falls above the unity line. This is in line with the observed complex morphology of the M\,31 nucleus, while the MW NSC structure and SB-profile are undisturbed by the presence of the MBH in its center. In other words, the large size ratio relates to the larger fraction of the M\,31 NSC stars (mass) within the \smbhsize\ that are affected by the MBH \cite[as also discussed in e.g.][]{Peiris&Tremaine03}. We note that the majority of the galaxies in the \cite{Neumayer&Walcher12} sample have a size ratio below one. It may be interesting to check whether those galaxies with size ratios similar or greater to M\,31 have similarly complex central morphologies. 

Another question to ask from this observational perspective that can be related to theoretical expectations is to what extent galaxies with a SMBH could also harbor a ``classical`` NSC with a radius in the range 2\,-\,5\,pc? To address this question for the \cite{McConnell&Ma13} sample of ``pure'' MBHs, we show in Figure\,\ref{fig:Gal_mass_rBH_toNSC}\,b the ratio between the derived \smbhsize\ and a ``nominal'' NSC size of \nscsize=3\,pc, plotted against host galaxy mass\footnote{We assume a value of 3\,pc because it is the most representative (typical) value for \nscsize\ \cite[cf. Fig.\,11 in][]{Georgiev&Boeker14}.}. The vast majority of the systems have a size ratio that falls significantly above unity, which is in line with theoretical expectations for the absence of a NSC.

On the other hand, one could also assume that a putative NSC in these galaxies has a mass corresponding to the extrapolation of the \nscsize - ${\cal M}_{\rm\star, gal}$ relation shown in Figure\,\ref{fig:Galaxy_NSCs_reff}. In this case, it implies that the NSC initially outgrew the MBH by a large amount, the theoretical size ratios are significantly smaller (see Figure\,\ref{fig:Gal_mass_rBH_toNSC}\,), but still fall above unity, again favouring the strong impact by the SMBH on the NSC structure and its stellar velocity field. We further discuss these observations and their implications in \S\,\ref{Sect:Discussion:CMO}.

\section{Discussion}\label{Sect:Discussion}

We have found noticeable differences in the fitted relations between \nscmass\ / \nscsize\ and host galaxy mass for different morphological types, as shown in Figures\,\ref{fig:NSC_mass_size} and \ref{fig:Galaxy_NSCs_reff}.
We now investigate whether these differences could provide insight into
the evolutionary path of NSCs in different hosts. In other words, can the properties of NSCs (size, mass) be traced to the various growth mechanism(s) that result from the internal secular evolution of the host, such as gas accretion and/or merging star clusters? For example, in late-type galaxies that are gas rich and harbor young stars and star clusters, NSC growth is more likely to be ongoing, while in early-type hosts, the only feasible mechanism today is the infall of old stellar populations.

\subsection{Possible measurement biases in \nscsize\ and \nscmass }\label{Sect:bias}

We first discuss possible biases in the estimates for the sizes and photometric masses of NSCs. In late-type galaxies, the derived values for \nscmass\ can be affected if the light (and hence the color) of the NSC is significantly influenced by a young stellar population (cf. \S\,\ref{Sect:Data_Gal_Mags}). This effect can cause the $M/L$ ratio (and thus \nscmass ) to be underestimated by up to a factor of 5, for example if 10\% of the stellar mass is $\approx$5\,Gyr younger than the rest (cf. \S\,\ref{Sect:Data_NSCs}). However, this is opposite to what is observed in Figure\,\ref{fig:NSC_mass_size}, namely that at fixed \nscsize\,
NSCs in late-type galaxies are \emph{more} massive than those in early-type hosts. We conclude that the actual offset between the two populations in Figure\,\ref{fig:NSC_mass_size} may well be more pronounced.  

Another possible bias comes from underestimating \nscsize\ in late-type hosts if the NSC contains a significant fraction of young stars that are more centrally concentrated. We detected such an effect in \cite{Georgiev&Boeker14} by measuring the ratio of NSC sizes in blue and red passbands (see also \citeauthor{Kormendy&McClure93} \citeyear{Kormendy&McClure93}, \citeauthor{Matthews99} \citeyear{Matthews99} and \citeauthor{Carson15} \citeyear{Carson15}). However, as shown in Figure\,10 of \cite{Georgiev&Boeker14}, this bias is $<5$\% for our NSC sample, and is thus a negligible effect when interpreting Figure\,\ref{fig:NSC_mass_size}. We also do not expect a significant measurement bias caused by any contamination of NSC light from the underlying disk and/or bulge, because i) the galaxies in our late-type sample are selected to have a low inclination (see \S\,\ref{Sect:Data_NSCs}) which minimizes this effect, and ii) our PSF-fitting methods implicitly account for any ''background`` emission surrounding the NSC. Within 0.2\arcsec, only a very steep bulge would be of a concern, however, by construction of our catalogue of very late-types, we have no such cases.

A last possible bias in measuring \nscsize\ and \nscmass\ in early-type hosts may arise from an imperfect decomposition of the combined NSC-bulge surface brightness profile. This effect is more pronounced for luminous bulges with steeply rising surface brightness profiles. \cite{Cote06} tested how well \nscsize\ and $M_V$ can be recovered by generating simulated data of NSCs with a range in size and luminosity.  They find that irrespective of the input NSC size, \nscsize\ is recovered to better than 15\%, with a bias toward underestimating \nscsize\ with increasing NSC magnitude. Accounting for such a bias would \emph{increase} the offset between the early- and late-type samples in Figures\,\ref{fig:NSC_mass_size} and \,\ref{fig:Galaxy_NSCs_reff}. As for the inferred luminosity (i.e. mass) of the NSC, \cite{Cote06} estimate that it can be overestimated by $<0.1$\,mag for bright NSCs, and by as much as 0.5\,mag for the faintest NSCs. This means that NSC masses in early-type hosts may be overestimated by up to a factor of three - again causing the separation of the two subsamples to become more pronounced.

We thus conclude that the differences between the early- and late-type samples seen in Figures\,\ref{fig:NSC_mass_size} and \ref{fig:Galaxy_NSCs_reff} cannot be explained by observational biases in deriving \nscsize\ and \nscmass , and in fact are likely to be more pronounced when observational biases are fully accounted for. This strengthens our finding that NSCs in late-type galaxies are more compact, both at fixed NSC mass and at fixed host galaxy mass.

\subsection{Differences in NSC properties for different host morphologies}\label{Sect:Discussion:differences}

As discussed in the last section, measurement biases are insufficient to explain the result that NSCs in late-type galaxies are more compact, both at fixed NSC mass and fixed host galaxy mass (Figures\,\ref{fig:NSC_mass_size} and \ref{fig:Galaxy_NSCs_reff}). An obvious question to ask therefore is which, if any, evolutionary effects could explain this difference? An increase in NSC size and mass over time has been demonstrated by a number of numerical simulations of merging clusters \citep{Fellhauer&Kroupa02a,Baumgardt03,Bekki04,Bruens11,Antonini13,Gnedin14,ArcaDolcetta14} or/and mass build up from gas accretion \citep{Hartmann11}. In particular, the slope of the \nscsize--\nscmass\ relation found in this paper ($\alpha\simeq0.34^{+0.05}_{-0.04}$, cf Table\,\ref{Table:fits}) is consistent with the slope of 0.4 found from cluster merger simulations \citep[e.g.][]{Bekki04}. This suggests that the smaller sizes of NSCs in late-types may well be explained by a scenario in which late-type nuclei have not (yet) experienced the infall of a large number of stellar clusters, i.e. that they are ``lagging behind'' their counterparts in early-type hosts in the accumulation of stellar mass. Moreover, in late-type galaxies in-situ star formation may be the driving mechanism to grow the NSC, leading to a higher phase-space density and thus smaller sizes than what can be reached by cluster merging.

What can the difference between late- and early-types in the \nscmass--${\cal M}_{\rm\star, gal}$ relation in Figure~\ref{fig:Gal_NSCs_mass}, where early-type nuclei show a steeper slope, tell us in this context? As discussed above, \nscmass\ in early-types may well be overestimated by about a factor of three in the most luminous host galaxies. While this effect certainly contributes to the steeper slope of early-type nuclei, there are also plausible evolutionary effects that may explain this. For example, the more eventful merger history of massive early-type hosts likely leads to an over-proportional growth of their NSCs caused by enhanced funneling of material to the center, both in the form of gas and star clusters.
Late-type hosts, in contrast, have not experienced significant mergers, and in this scenario, their nuclei would grow only proportionally to their host mass, resulting in a shallower slope compared to early-types.

As illustrated in Figure\,\ref{fig:Gal_NSCs_mass_all}\,b, disk-dominated NSC host galaxies contain significant amounts of (HI) gas. In a scenario in which late-type galaxies eventually turn into early-type galaxies, one can ask how their NSCs move from the steeper late-type relation in Figure\,\ref{fig:Gal_NSCs_mass_all}\,b to the shallower early-type relation. One possible path is to simply remove the gas, e.g. by ram pressure stripping, and galaxy-galaxy ``harassment'' in a cluster environment. For a galaxy in isolation, the only plausible path to remove significant amounts of gas are stellar winds and/or supernovae from starburst regions \citep[e.g.][]{MacLow&McCray88,Meurer95}. 
In fact, some low-mass late-type galaxies exhibit wind velocities above 1000\,km/s, while their escape velocity is only $400-500$\,km/s \cite[e.g.][]{Strickland&Heckman09}. Such winds will naturally have the largest impact on the lowest mass galaxies due to their weaker potentials \cite[e.g.][and refs. therein]{Carraro14}. This could offer an evolutionary path for a NSC host galaxy from one relation to the other in Figure\,\ref{fig:Gal_NSCs_mass_all}\,b, even without involving interactions, leading to a shallower \nscmass-${\cal M}_\star$ relation when the HI gas mass is removed. However, due to the galaxy density environments dichotomy between our samples, we can not exclude the possibility that environmental effects could make inapplicable one or the other discussed effects for cluster or isolated galaxies.

We conclude that the differences in \nscsize\ between late- and early-type galaxies are likely due to NSCs and their host galaxies being at different evolutionary stages. The differences in \nscmass\ between the two morphological host types could possibly be explained by measurement biases, however, some plausible evolutionary effects can not be ruled out.

\subsection{Relations between NSCs and MBHs}\label{Sect:Discussion:CMO}

In this section we discuss what the relations between host galaxy stellar mass ${\cal M}_{\star,\rm gal}$ and the parameters ${\cal M}_{\rm MBH + NSC}$, \bhmass/\nscmass\ and \smbhsize/\nscsize\ (\S\,\ref{Sect:SMBH_NSC_coexistence}) can tell us from observational point of view about the interplay between these two types of object.

The apparent lack of systems with MBH at galaxy masses below $10^9\,M_\odot$ is noteworthy. The extrapolation of the ${\cal M}_{\rm NSC+BH}$-${\cal M}_{\star\rm gal}$ relation for coexisting NSCs and MBHs towards lower galaxy masses implies that in this range, a central MBH is expected to have a mass of \bhmass$\lesssim10^4-10^5M_\odot$. At typical galaxy distances of a few Mpc or more, this is below the detection limit of current instruments and analysis techniques.

On the other hand, there is evidence that low- to intermediate-mass BHs  reside in (some) massive globular clusters (e.g. \citeauthor{Luetzgendorf13}\,\citeyear{Luetzgendorf13}, but see also \citeauthor{Lanzoni13}\,\citeyear{Lanzoni13}). They appear to define a shallower scaling relation, which can potentially be explained if they are the remnant nuclei of stripped galaxies. Indeed, this is a popular formation scenario for dense stellar systems with MBHs, such as UCDs \citep{Mieske13,Seth14}. 

At the high mass end, on the other hand, NSCs appear to become rare. This likely implies that as a galaxy grows in mass, there are processes at work that destroy NSCs, or prevent their formation on the first place. As discussed in \S\ref{Sect:BH.NSC_Sizeratio}, the ``classical'' NSC surface brightness profile, which is normally well described by a King model, may no longer be a good representation if the radius of the MBH sphere of influence (\smbhsize) is significantly larger than the NSC effective (or half-mass) radius. In cases where there is only a MBH in the nucleus, the dissolution of an infalling NSC that passes through \smbhsize\ has been demonstrated in simulations \cite[][]{Antonini13,Mastrobuono-Battisti14}. 
A similar situation can also arise if a MBH spirals into the nucleus occupied by a NSC (e.g. in an Antennae-like merger where two galactic nuclei will coalesce; see \cite{Antonini15} for the effect of mergers).  

In the absence of infalling external objects, it is less clear how an NSC could be destroyed. As discussed in \S\ref{Sect:BH.NSC_Sizeratio}, in cases where NSC and MBH coexist, i.e. in the intermediate galaxy mass range, the NSC would be destroyed if it is ``outgrown'' by the MBH. A potential example for this process is M\,31 where the strong dynamical impact of the MBH on its surroundings are clearly present \cite[e.g.][]{Peiris&Tremaine03}. The inner few\,pc of the M\,31 nucleus is strongly axisymmetric and composed of three main central components - a central blue component at $\lesssim$ 0.2 pc along with a double-lobed redder component at $\sim$ 1-2 pc (i.e. each lobe is located on either side of the central blue component), which can be explained as the projection of an edge-on central disc of stars on Keplerian orbits around the MBH \cite[][]{Tremaine95,Peiris&Tremaine03,Brown&Magorrian13}. 

On the other hand, in the Milky Way nucleus, the other well-studied example of NSC-MBH system, the NSC has retained a normal star cluster profile that is well described by a standard King model \citep{Schoedel14}. This is expected due to the fact that, in contrast to M\,31, the \smbhsize/\nscsize\ (or \bhmass/\nscmass) ratio in the Milky Way is less than 1 (cf. Figs.\,\ref{fig:MBH_NSC}b and \ref{fig:Gal_mass_rBH_toNSC}a). Next generation of large telescopes and instrumentation will help to extend this type of comparison to other nuclei with coexisting NSC and MBH. As pointed out by \cite{Georgiev&Boeker14}, there are a number of NSCs that are poorly described by a King model, and constraining their \bhmass/\nscmass\ (or \smbhsize/\nscsize) ratios would allow to check whether internal evolution due to the presence of a MBH is a viable explanation for their complex morphologies.

Of course, it cannot be ruled out that the mass ratio \bhmass/\nscmass\ is not affected by evolutionary effects at all, but is instead governed by the inability to form either object in the first place due to destructive feedback from the formation of the other. This ``competitive feedback'' scenario has been discussed by \cite{Nayakshin09}. 

\section{Conclusions}\label{Sect:Conclussions}

We presented an updated analysis of various scaling relations between Nuclear Star Cluster (NSC) properties (mass and size) and the stellar mass of their host galaxies. We compared these scaling relations between late- and early-type host galaxies, aided by the recent compilation of NSC properties in a large sample of late-type galaxies \citep{Georgiev&Boeker14}. We added literature estimates of NSC properties in a number of other late- and early-type galaxies. Of special relevance are data for NSCs that harbor a massive black hole (MBH). 

Our study expands on earlier works \citep{Seth08b,Erwin&Gadotti12} that consider the total galaxy stellar mass (bulge plus disk), instead of only the bulge mass. This is especially relevant for late-type hosts which have most of their mass in the disk and therefore provides a more complete picture of the potential supply of matter (e.g. gas and star clusters) to the nucleus. 
For comparative purposes we add to the baryonic mass budget in these galaxies, their HI and X-ray masses (\S\,\ref{Sect:HI_Xray_Masses}) to illustrate the amount of available material to further supply the evolution of the CMO.

We summarize our main results and their implications as follows:
\begin{itemize}
\item We provide photometric masses for all NSCs as well as their host galaxies, calculated from color-dependent mass-to-light ratios. These masses are listed in Table\,\ref{Table:GB_Gal_NSC} (full version is available in the online version).

\item The NSCs have a typical mass of a few\,$\times10^6M_\odot$ and constitute \nscmass/${\cal M}_{\star\rm gal}\simeq0.1\%\pm0.2\%$ of the total galaxy stellar mass (\S\,\ref{Sect:NSC_gal_relations}, Fig.\ref{fig:Gal_NSCs_mass},\,\ref{fig:Gal_NSCs_mass_all}), consistent with earlier results.

\item We derive empirical scaling relations between \nscsize\ and \nscmass\ and host galaxy total stellar mass for NSCs in late- and early-type host galaxies. The fit values of these scaling relations are provided in Table\,\ref{Table:fits}.

\item The mass--size relation for NSCs shows a $\sim1.5\sigma$ significant difference between late- and early-type galaxies (Fig.\,\ref{fig:NSC_mass_size}) that cannot be explained by plausible measurement biases. At a given \nscmass , NSCs in late-type hosts are on average twice as compact as their counterparts in early-type hosts (\S\,\ref{Sect:NSC_mass_size_relation}). We interpret this as evidence that NSCs in late-type galaxies are still evolving, i.e. they still have growth potential via gas accretion and/or cluster merging in the nucleus.

\item The \nscmass\,-\,${\cal M}_{\star\rm gal}$ scaling relation for NSCs in early-type hosts has a steeper slope than that for NSCs in late-type galaxies. Specifically, NSCs in early-types become progressively more massive with increasing total galaxy mass, compared to NSCs in late-type galaxies (Fig.\,\ref{fig:Gal_NSCs_mass}\,c. We interpret this result as likely being due to measurement bias, which can reach a factor of three in \nscmass\ in massive early-type galaxies. However, we can not exclude the possibility that the difference in slopes is real, as a number of physical processes could contribute to this trend, such as i) depletion of the host galaxy mass via ram pressure stripping and/or galaxy ``harassment'', or ii) accelerated NSC growth in massive hosts due to their enhanced merger history.

\item Coexisting NSC-MBH systems define a ${\cal M}_{\rm BH+NSC}$\,-\,${\cal M}_{\star\rm gal}$ relation (\S\,\ref{Sect:SMBH_NSC_coexistence}, Fig.\,\ref{fig:MBH_NSC}) with a slope consistent with that defined by NSCs without MBHs in early-types, but steeper than both the well-known \bhmass\,-\,${\cal M}_{\rm bulge}$ relation and the relation defined by late-type NSCs without MBHs. To within the fit uncertainties, the slopes of the \nscmass\,-\,${\cal M}_{\star\rm gal}$, ${\cal M}_{\rm BH+NSC}$\,-\,${\cal M}_{\star\rm gal}$ and \bhmass\,-\,${\cal M}_{\rm bulge}$ relations are consistent with each other. This is probably suggesting similar physical  mechanisms driving NSC or/and MBH growth as a function of galaxy mass.

\item We looked at the size ratio between the MBH sphere of influence and the NSC effective (or half-mass) radius. It covers a wide range of values ($0.01 \leq$ \smbhsize/\nscsize $\leq 100$, \S\,\ref{Sect:BH.NSC_Sizeratio}, Fig.\,\ref{fig:Gal_mass_rBH_toNSC}), and because \smbhsize/\nscsize $\equiv$ \bhmass/\nscmass, the limit \smbhsize/\nscsize $\geq 1$ implies that more than 50\% of the bound NSC stars are on Keplerian orbits around the MBH. The best example for this scenario is the nucleus of M\,31, which has a \smbhsize/\nscsize$>\!1$, thus illustrating the dynamical influence of the MBH on its surroundings. The NSC-MBH system in the Milky Way nucleus, in contrast, has a size ratio below 1. It has thus, unsurprisingly, a surface brightness profile that is well described by a King model.

\end{itemize}

\section*{Acknowledgments}

IG would like to thank the science department of ESA-ESTEC in Noordwijk and the ESO-Garching visitor programme for partial support during the preparation of this paper, as well as the Max-Plank-Institut f\"ur Astronomie (Heidelberg) for support during the final stages of this work. NL gratefully acknowledges the generous support of an NSERC PDF award. We also thank Morgan Fouesneau for numerous discussions and help with fitting techniques.\\
This research has made use of the NASA/IPAC Extragalactic Database (NED) which is operated by the Jet Propulsion Laboratory, California Institute of Technology, under contract with the National Aeronautics and Space Administration. We also acknowledge use of the HyperLeda database (http://leda.univ-lyon1.fr).

\bibliographystyle{aa}
\bibliography{references}

\clearpage
\newpage
\appendix

\section{Fitting method basics}\label{Sect:FitDetails}

There are several methods to fit a straight line through data. Only few treat measurement uncertainties and finite data samples, both of which are important for inferring the most likely values and uncertainties \cite[e.g.][]{Hogg10,Mengersen13,Zhu15}. The method adopted here for fitting the various subsamples with a straight line (in a $\log-\log$ space) also treats non symmetric uncertainties. It consists of a maximum likelihood estimation (MLE) of the model parameters and data bootstrapping to account for the finite data samples and build the probability density distribution (posterior) of model parameters \citep{Nelder&Wedderburn72,Hogg10}. This method is very similar to an MCMC, however, instead of sampling the prior space to derive posterior density distributions, we sample from the bootstrapped data. The functional form of the measurement uncertainties used for the bootstrapping as well as one part of the noise model in the MLE is expressed as a combination of two Gaussians joined by common mode value, which in our case is the location ($\mu$) of the data point:\\
$f(t;\mu,\epsilon_1,\epsilon_2)= A \exp (- \frac {(t-\mu)^2}{2 \epsilon_1^2})$, if  $t < \mu$, and\\
$f(t;\mu,\epsilon_1,\epsilon_2)= A \exp (- \frac {(t-\mu)^2}{2 \epsilon_2^2})$ otherwise,\\
where $A = \sqrt{2/\pi} (\epsilon_1+\epsilon_2)^{-1}$, and $\epsilon_1,\epsilon_2$ is each side of the uncertainty. This distribution is known as a split normal distribution \citep{Gibbons&Mylroie73,John82}. The product of these functions along $x$ and $y$, $F\!=\!f(x,\mu_x,\epsilon_{x1},\epsilon_{x2})\times f(y,\mu_y,\epsilon_{y1},\epsilon_{y2})$ allows to treat each data point with a probability density space defined by a split normal distribution, $\epsilon_{xy}\!=\!\{\epsilon_{x1},\epsilon_{x2},\epsilon_{y1},\epsilon_{y2}\}$.

Our dataset can be defined as, $D$ of $k$ independent observations, $d_k\!=\!\{x_k,y_k\}$ with uncertainties $\epsilon\!=\!\{\epsilon_k^2\}\!=\!\{\epsilon_{x1}^2,\epsilon_{x2}^2,\epsilon_{y1}^2,\epsilon_{y2}^2\}$ described by the split normal distribution. The model, $M$ (straight line with parameters, $\{\alpha,\beta\}$) is ``predicting'' how the data should be distributed. We therefore do not test for other than linear (in log-log space) relation between the fitted quantities. The model dispersion is also described by a Gaussian model ($\Sigma$) with a variance ($\sigma^2$) orthogonal to the linear regression. The ``noise'' we can also note as $E$ ($\{\epsilon,\Sigma\}$). Thus, the model can be written as:\\
$y_k=\alpha\times x_k +\beta +\epsilon+\Sigma$,\\[.1cm]
where $\alpha$ and $\beta$ are the slope and intercept model parameters. Thus, the density distribution of the data given the model can be expressed as:\\[.1cm]
$p(d_k|M,E)=\frac{1}{\sqrt{2\pi\sigma^2}}\exp\lbrace-\frac{1}{2\sigma^2}(y_k-\alpha\times x_k-\beta)^2\rbrace\times$\\[.1cm]
\hspace*{2cm}$\times f(x_k,\mu_{x_{k}},\epsilon_{x_{k}})\times f(y_k,\mu_{y_{k}},\epsilon_{y_{k}})$.\\[.1cm]
Following the Bayes rule, which states that the posterior probability distribution of ``observing'' a model $M$ (the linear regression) given the distribution of the data ($D$) and its uncertainties ($E$, incl. model noise) is:\\
$P(M|D,E)=\prod\limits_{k=1}^{N} p(d_k|M,E)\times\frac{P(M,E)}{P(D,E)}$,\\
where $P(D,E)$ is the evidence, i.e. the probability of the data averaged over all parameters (it also assures that the posterior distribution integrates to unity), $p(d_k|M,E)$ is the above likelihood of the $k-$th data point given the model $M$. Since the denominator (the data) does not depend on the model parameters, $\{\alpha,\beta\}$, the Bayesian estimator is obtained by maximizing the likelihood $p(d_k|M,E)P(M,E)$ with respect to the model (straight line) parameters. Assuming that the model parameters $\alpha{\ \rm and\ }\beta$ are uniformly distributed, then the Bayesian estimator is obtained by maximizing the likelihood function $p(d_k|M,E)$. Or for convenience, to convert products into to sums, the natural logarithm of it:\\[.1cm]
$\ln\ p(d_k|M,E)=-\frac{N\ln2\pi}{2} -\frac{1}{2}\!\sum\limits_{k=1}^{N}\big[ \ln\sigma^2-\frac{(y_k-\alpha x_k-\beta)^2}{\sigma^2}-$\\[.1cm]
\hspace*{3.8cm}$-\ln F(x_k,y_k,\mu_{x_{k}},\mu_{y_{k}},\epsilon_{x_{k}},\epsilon_{y_{k}})\big]$\\[.1cm]

In general, this treatment, that followed from discussion in \cite{Bailer-Jones15} will be described in details in Fouesneau et al.\,(in prep.). For normal distributions the likelihood function has a closed form expression for the estimator, however, if the estimator lacks a closed form, a solution can be obtained by MCMC. The estimator in our case has a closed form, however, we build our posterior distributions not by sampling from a wide model prior space \cite[with MCMC technique, e.g. {\it emcee}][]{Foreman-Mackey13}, but by sampling from a smaller data ``prior'' space restricted by the data uncertainties. This data space is generated by bootstrap from the split normal distribution of the measurement uncertainties. We note that the classical bootstrapping ($N$ new data points) makes significant difference only for small or/and data with large scatter, $\sigma^2$. In other words, the final posterior distribution describing the posterior distributions from $j-$ number of model estimators is defined by:\\
$P(M|D,E)=\sum\limits_{j=1}^{n}\prod\limits_{k=1}^{N}P(d_{k,j}|M_j,E_j)\times P(M_j)$,\\
where $n$ is the number of new data samples (bootstraps). This number in our case was chosen to be 1500 to sample well the posterior distributions. The linear regression model fitted to our data is of the form:
\begin{equation}
\log_{10}(y/c1)=\alpha\times\log_{10}(x/c2)+\beta,
\end{equation}
where $c1,c2$ are normalization constants, $\alpha$ and $\beta$ are the slope and intercept, respectively. As also discussed in \S\,\ref{Sect:NSC_mass_size_relation}, the right choice of normalization constants is important to minimize the correlation between the slope and intercept. This provides a more realistic estimation of the uncertainties for $\alpha,\beta$ and the dispersion, $\sigma$, of the data, because it determines the shape of the posterior probability density functions (PDFs). Those distributions are shown by histograms in Figure\,\ref{fig:A.fits}. The $c1,c2$ constants are estimated a priori from the highest probability density value of the 2D-PDFs of the data. Figure\,\ref{fig:A.fits} illustrates the results from fitting each parameter pair. Each line in the top-right panel represents one of the possible solutions (one bootstrap realisation of the data described by the uncertainty of each data point) fitted with linear regression, which coefficients are shown with dots in the lower-left panel. For illustration, contour lines show the density distribution of the probable solutions for the slope and intercept. The PDFs of $\alpha,\beta$ are shown with histograms, where with red lines is indicated their highest probability modal value. The solid red line in the top-right panel shows the final solution for the respective subsample. The two red, parallel lines indicate the $1\sigma$ dispersion of the data with respect the solid line of the final solution.

\begin{figure*}
\subfloat{\includegraphics[width=0.5\textwidth, bb=30 20 566 392]{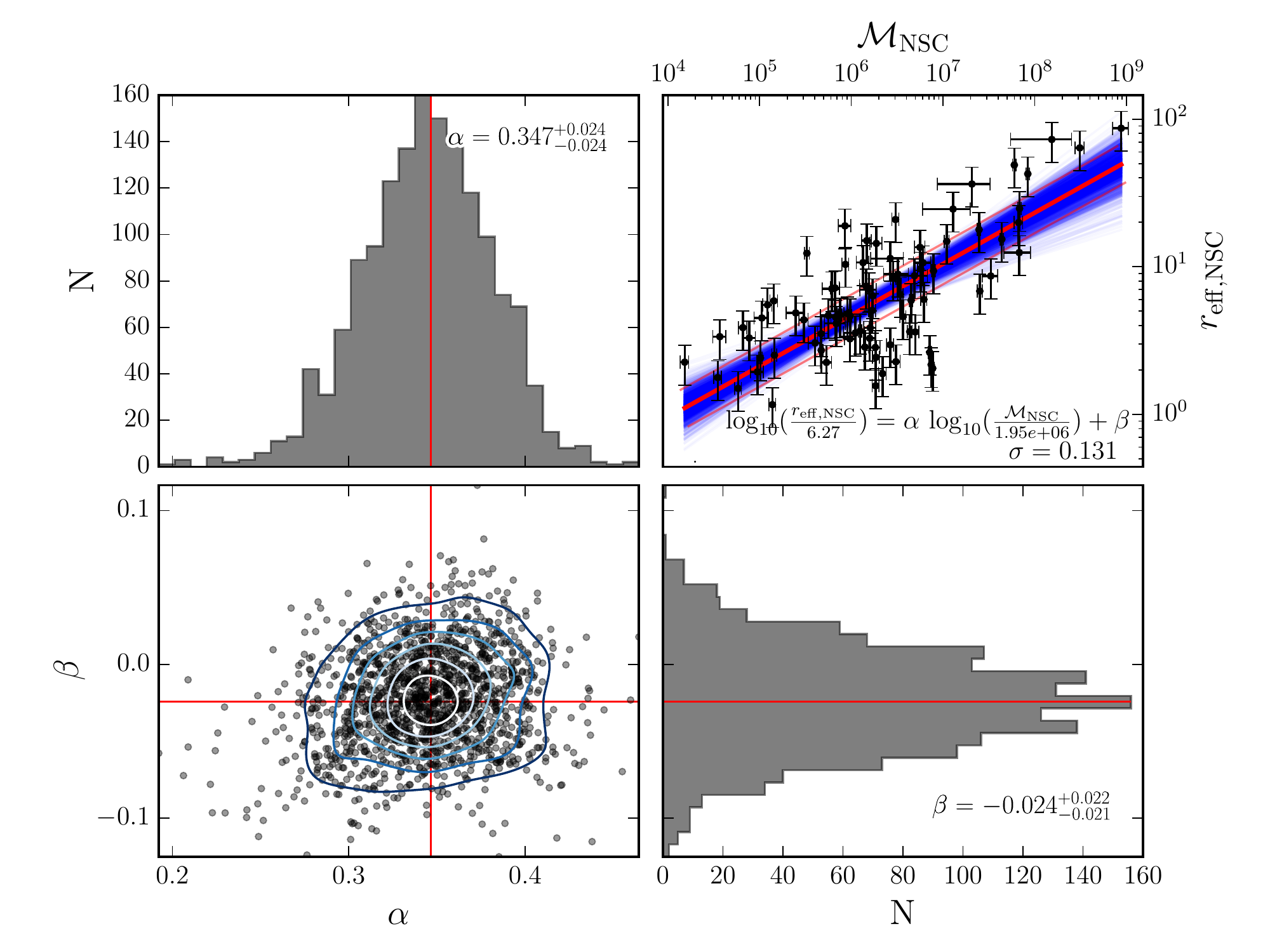}}
\subfloat{\hspace*{.38cm}\includegraphics[width=0.5\textwidth, bb=30 20 566 392]{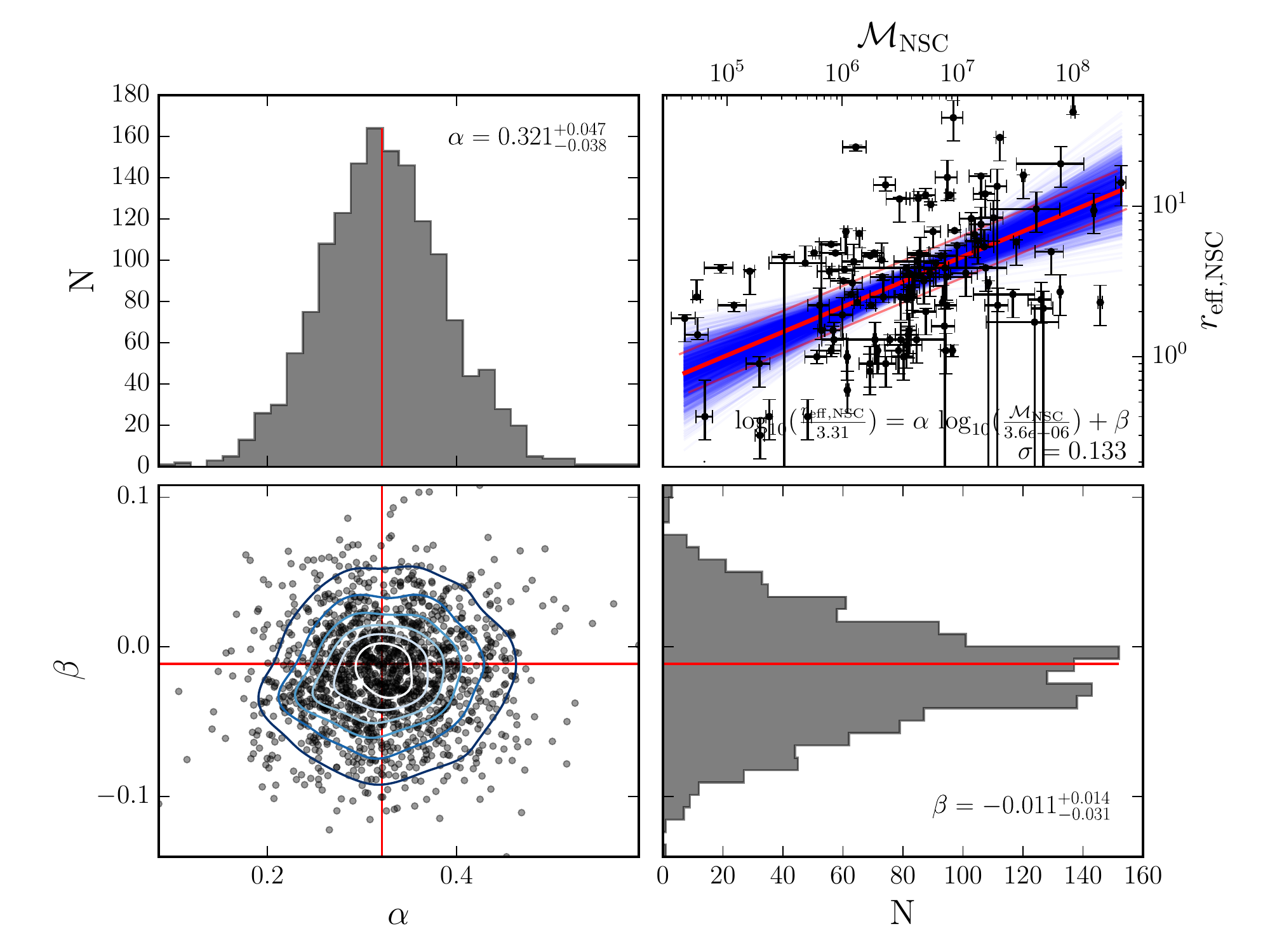}}
\\
\subfloat{\includegraphics[width=0.5\textwidth, bb=30 20 566 416]{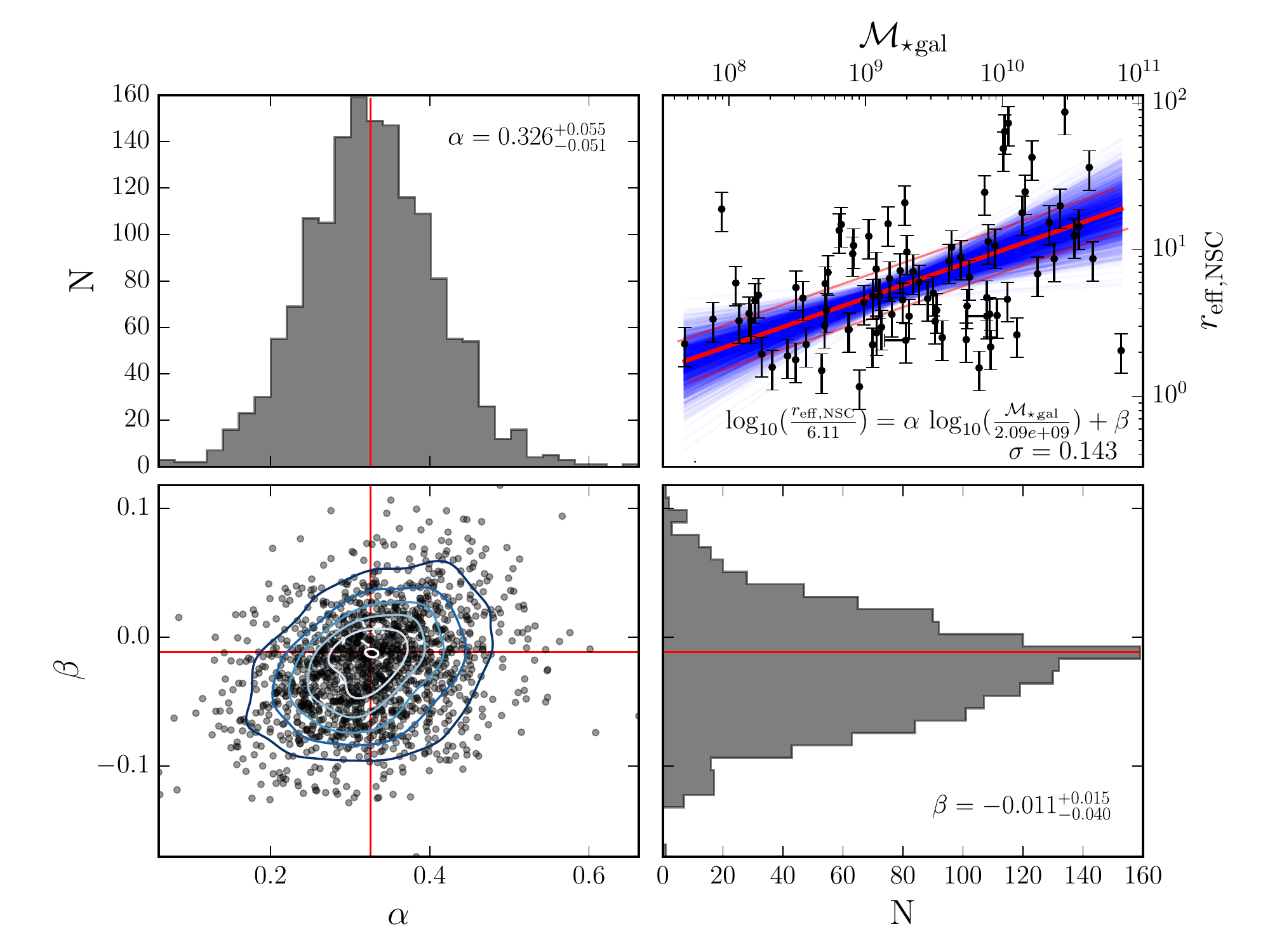}}
\subfloat{\hspace*{.38cm}\includegraphics[width=0.5\textwidth, bb=30 20 566 416]{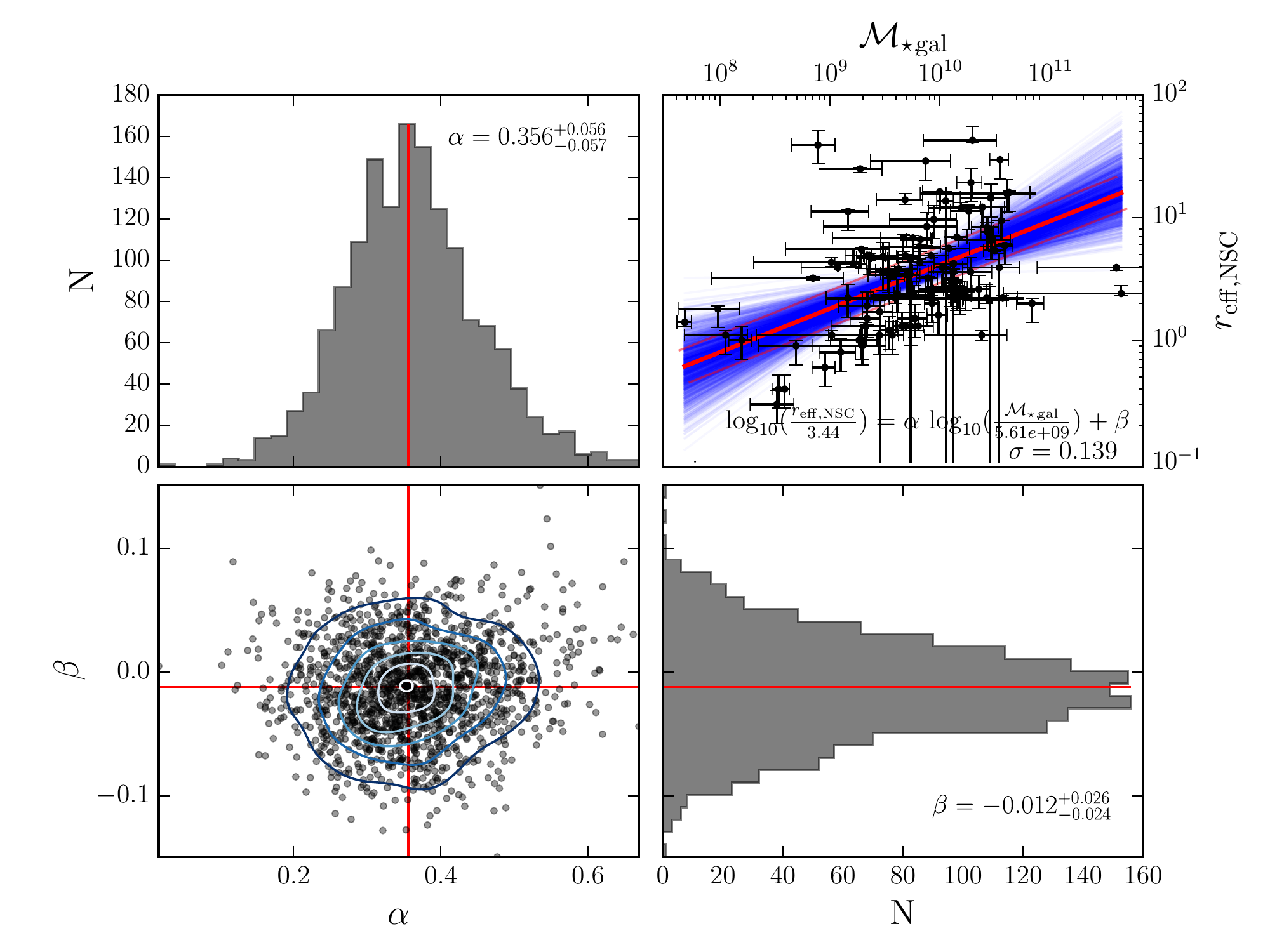}}
\\
\subfloat{\includegraphics[width=0.5\textwidth, bb=30 20 566 416]{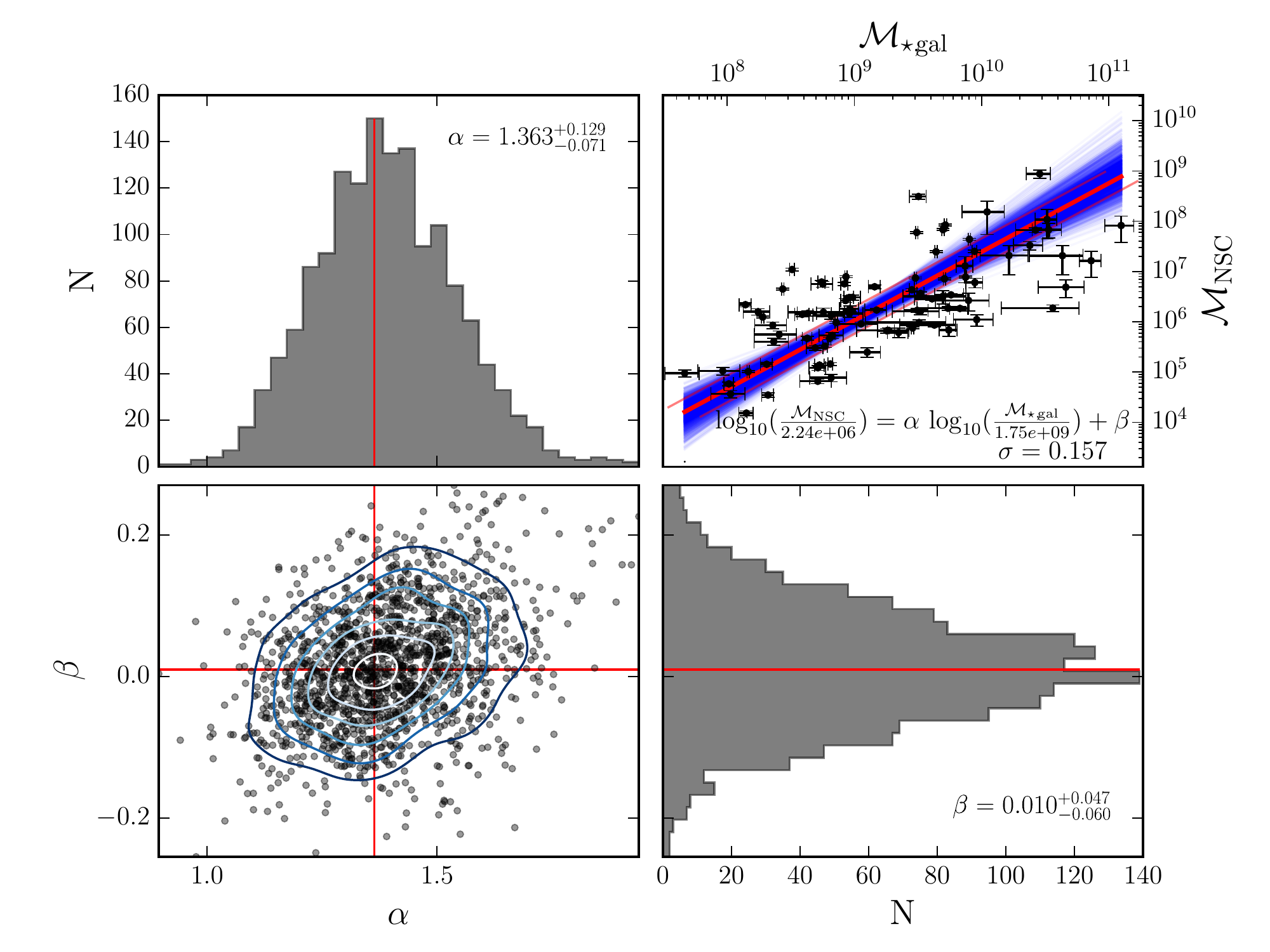}}
\subfloat{\hspace*{.38cm}\includegraphics[width=0.5\textwidth, bb=30 20 566 416]{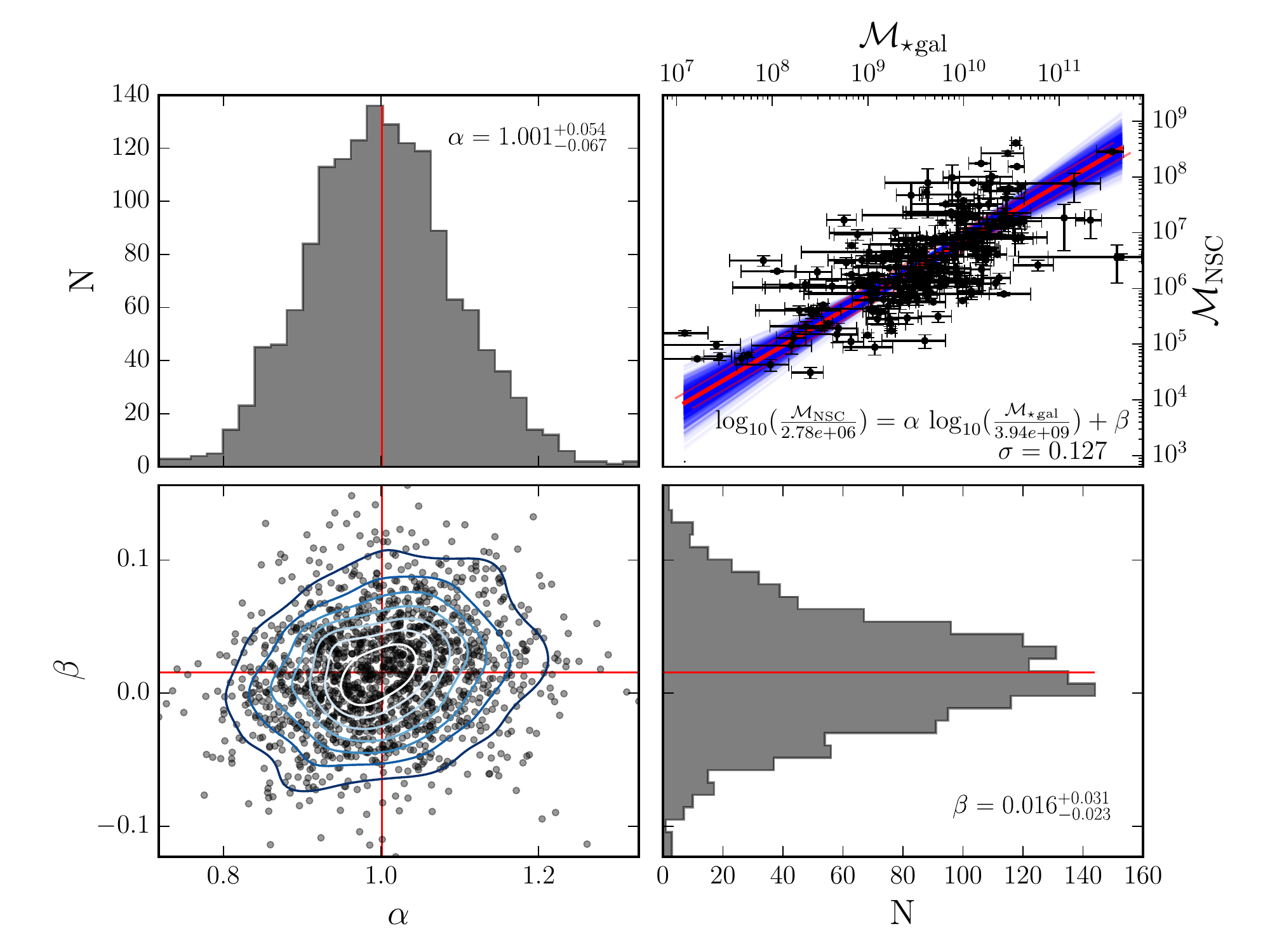}}
\caption{Results from the bootstrapped, non-symmetric error weighted, maximum likelihood fitting technique. {\bf Left} and {\bf Right} figure columns are for early- and late-type galaxies, respectively. {\bf Top row:} NSC effective radius versus NSC mass (Fig.\,\ref{fig:NSC_mass_size}); {\bf middle row} NSC effective radius versus host galaxy stellar mass (Fig.\,\ref{fig:Galaxy_NSCs_reff}); {\bf bottom row:} NSC mass - host galaxy stellar mass (Fig.\,\ref{fig:Gal_NSCs_mass}). For each figure, the {\bf top-left} and {\bf bottom-right panels} show the projected distributions for each realization of the relation slope ($\alpha$) and intercept ($\beta$), which are shown with dots  in the {\bf bottom-left panel} and some of those are shown with lines in the {\bf top-right panel}. Their best value is estimated from the distribution mode value as indicated with solid lines (and density contours, in bottom-left). The thick (red) line plotted with the data ({\bf top-right panel}) shows the best solution, where the two parallel lines are the $1\!\sigma$ dispersion to that best fit.
}\label{fig:A.fits}
\end{figure*}

\newpage
\clearpage

\begin{figure*}
\ContinuedFloat
\subfloat{\includegraphics[width=0.5\textwidth, bb=30 20 566 392]{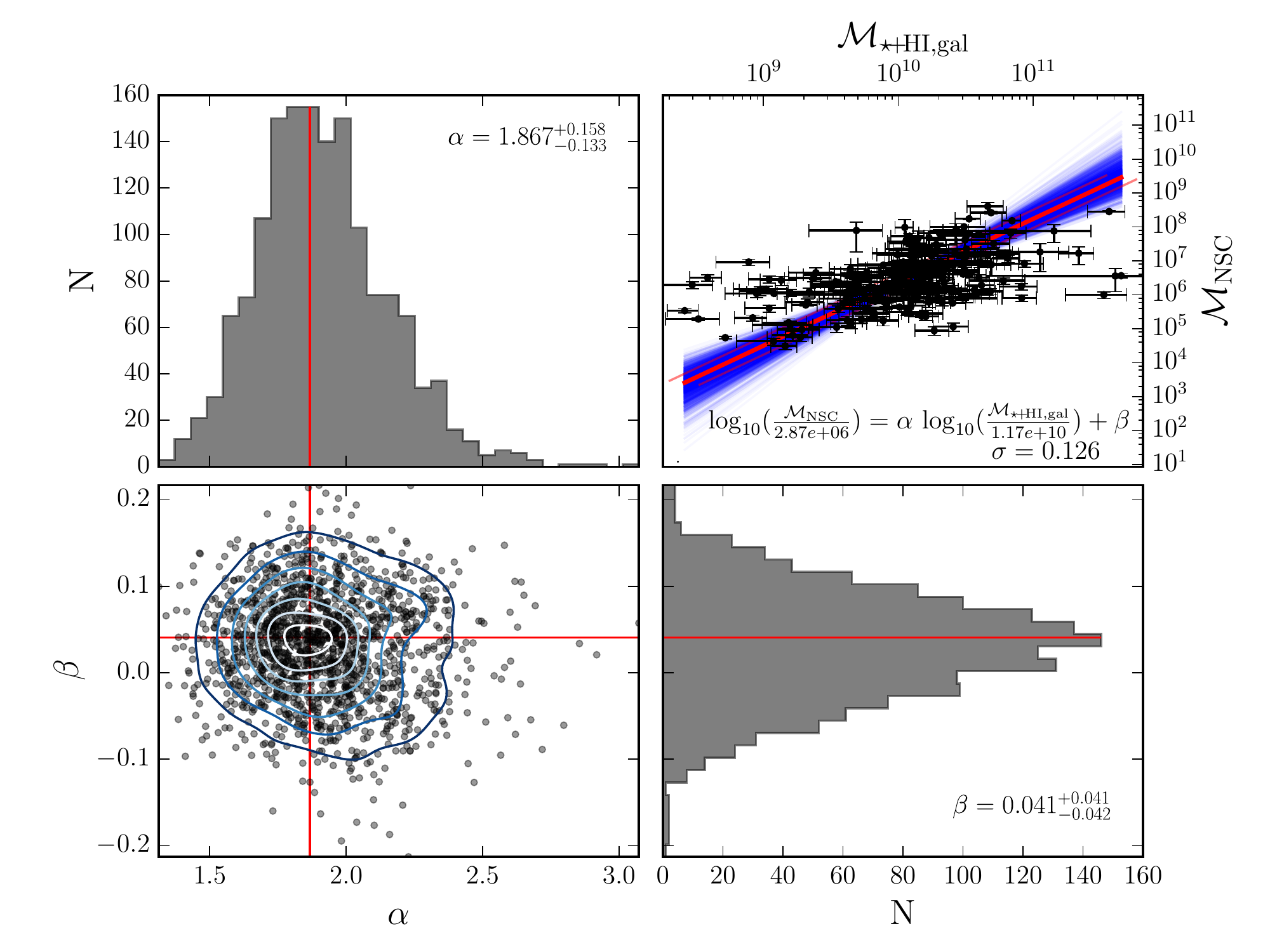}}
\subfloat{\hspace*{.38cm}\includegraphics[width=0.5\textwidth, bb=30 20 566 392]{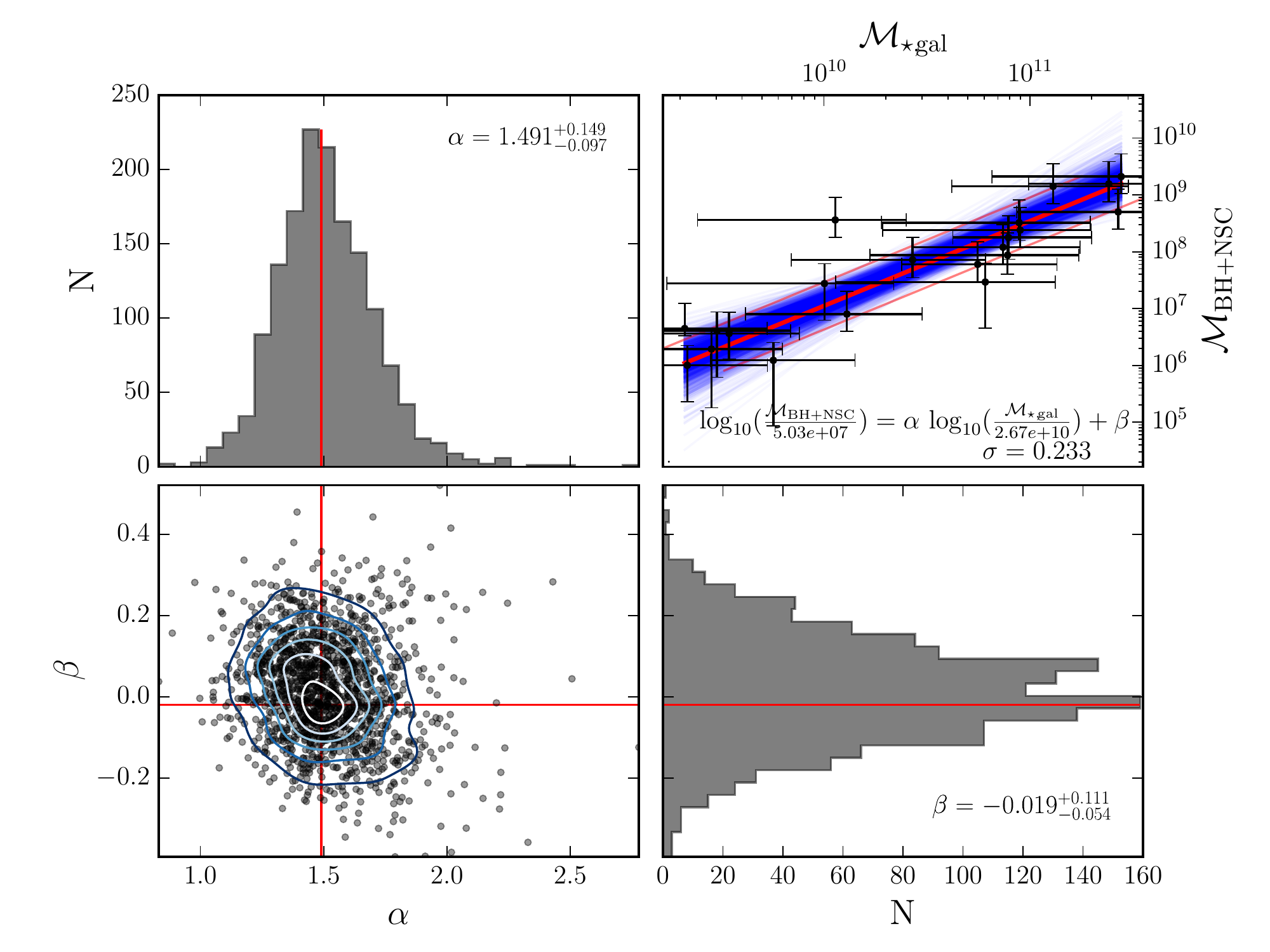}}
\caption{(\emph{Cont'd}) {\bf Left:} Fit for the NSC mass vs. host galaxy stellar plus HI mass, \nscmass$-{\cal M}_{\rm\star + HI, gal}$ (Fig.\,\ref{fig:Gal_NSCs_mass_all}\,b); {\bf Right:} Fit for the NSC plus MBH mass vs. host galaxy stellar mass, ${\cal M}_{\rm BH + NSC}-{\cal M}_{\rm\star, gal}$ (Fig.\,\ref{fig:MBH_NSC}).
}\label{fig:A.fits}
\end{figure*}

\newpage
\clearpage

\begin{deluxetable}{p{1.7cm}llllllllll}
\tabletypesize{\footnotesize}
%\rotate % For landscape mode
%\setlength{\tabcolsep}{0.15cm}
\tablecolumns{11}
\tablewidth{0pc} %%% <--- This is important!!! Otherwise it wont compile!!!!
\tablecaption{Photometric masses of Nuclear Star Clusters and their host galaxies. 
%NSC mass is calculated using photometric error weighted 
%average $M/L$s from the \protect\cite{BC03} SSP models for an assumed solar metallicity. NSC host galaxy mass is calculated using \protect\cite{Bell03} 
%$M/L-$colour relations. Details in \S\,\ref{Sect:Data_NSCs}, \ref{Sect:Data_Gal_Mags} and \ref{Sect:HI_Xray_Masses}.  
(Full table available online only)
\label{Table:GB_Gal_NSC}
}
\tablehead{

\colhead{Object} &
\colhead{$m\!-\!M$} &
\colhead{$t$} &
\colhead{\nscsize} &
\colhead{\nscmass} &
\colhead{${\cal M}_{\rm gal,\star}$} &
\colhead{${\cal M}_{\rm gal, HI}$} &
\colhead{NSC} &
\colhead{$M/L_{\rm NSC}$} &
\colhead{Galaxy} &
\colhead{$M/L_{\rm gal}$} \\
\colhead{} &
\colhead{[mag]} &
\colhead{} &
\colhead{[pc]} &
\colhead{[$10^4M_\odot$]} &
\colhead{[$10^7M_\odot$]} &
\colhead{[$10^7M_\odot$]} &
\colhead{filter} &
\colhead{[$M_\odot/L_\odot$]} &
\colhead{filter} &
\colhead{[$M_\odot/L_\odot$]} \\
\colhead{(1)} &
\colhead{(2)} &
\colhead{(3)} &
\colhead{(4)} &
\colhead{(5)} &
\colhead{(6)} &
\colhead{(7)} &
\colhead{(8)} &
\colhead{(9)} &
\colhead{(10)} &
\colhead{(11)}
}
\startdata

DDO078		&27.82     &10. 	   &$3.7_{-0.00}^{+0.1}$&   $19.21_{-0.48}^{+0.48}$&$	    1.22    \pm     0.91    $&$     0.00    \pm     0.00    $&$     VI      $&$      0.92    _{-     0.02    }^{+    0.02    }$&$     B       $&      1.16    \\
ESO138-G10     &30.84	  &7.9  	  &$7.7_{-7.7}^{+0.1}$&    $2427.8_{-60.69}^{+890.19}$&$   1824.23 \pm     743.8   $&$     3274.47 \pm     1432.55 $&$     I	   $&$      0.45    _{-     0.01    }^{+    0.17    }$&$     B       $&      1.16    \\
ESO187-G51     &31.13	  & 9.  	  &$1.8_{-0.8}^{+0.7}$&    $8.34_{-0.21}^{+3.06}$&$	   66.58   \pm     23.56   $&$     282.66  \pm     130.17  $&$     I	   $&$      0.45    _{-     0.01    }^{+    0.17    }$&$     BI      $&      1.64    \\
ESO202-G41     &  31.	  &8.9  	  &$1.8_{-1.8}^{+0.3}$&    $7.13_{-0.18}^{+2.62}$&$	   15.91   \pm     9.83    $&$     176.86  \pm     65.16   $&$     I	   $&$      0.45    _{-     0.01    }^{+    0.17    }$&$     BI      $&      0.75    \\
ESO241-G06     &31.44	  & 9.  	  &$2.2_{-0.4}^{+0.00}$&   $45.84_{-1.15}^{+16.81}$&$	   108.34  \pm     39.53   $&$     103.72  \pm     54.93   $&$     I	   $&$      0.45    _{-     0.01    }^{+    0.17    }$&$     BI      $&      1.43    \\
ESO271-G05     &32.57	  & 9.  	  &$3.1_{-0.6}^{+0.6}$&    $96.83_{-29.18}^{+32.62}$&$     104.11  \pm     105.00  $&$     114.43  \pm     50.06   $&$     V	   $&$      0.63    _{-     0.19    }^{+    0.21    }$&$     B       $&      1.16    \\
ESO290-G39     &31.33	  & 9.  	  &$4.6_{-4.5}^{+1.00}$&   $9.67_{-0.24}^{+3.54}$&$	   16.72   \pm     5.7     $&$     140.87  \pm     74.61   $&$     I	   $&$      0.45    _{-     0.01    }^{+    0.17    }$&$     BI      $&      0.63    \\
{\tiny ESO301-IG11}	&31.23     &9.9 	   &$3.7_{-0.4}^{+0.3}$&    $61.09_{-18.41}^{+20.58}$&$     410.68  \pm     151.75  $&$     0.00    \pm     0.00    $&$     V	    $&$      0.63    _{-     0.19    }^{+    0.21    }$&$     BI      $&      2.66    \\
\enddata

\vspace{-.5cm}
\tablecomments{
Columns: 
(1) NSC host galaxy ID, 
(2) Adopted distance modulus and 
(3) galaxy morphological code \cite[(1-2-3) are the same as in][]{Georgiev&Boeker14}; 
(4) NSC effective radius measured in $F606W$ or $F814W$ or $F450W$ or $F300W$, as available in this order. 
We note that differences in sizes in the different bands are $<10\%$ \protect\cite[see ][for discussion]{Georgiev&Boeker14}; 
(5) NSC mass calculated using the $M/L$ given in (9) for the $F606W$ (V) or $F814W$ (I) or $F450W$ (B) or $F300W$ (U) filter 
in this priority order, e.g. if Colour is $UBVI$, the value is the $M/L_V$ error weighted average from the $M/L_V$s from the 
various colour combinations containing $V$-band. NSC host galaxy stellar and HI mass (6 and 7), where the used colour and 
$M/L_B$ are in (10) and (11).
}

\end{deluxetable}

\label{lastpage}

\end{document}